\documentclass[aps,twocolumn,pra,10pt,nofootinbib]{revtex4-2}

\usepackage{multirow}
\usepackage{cancel}
\usepackage[normalem]{ulem}
\usepackage{dsfont}
\usepackage{amsmath}
\usepackage{bm}
\usepackage{amssymb}
\usepackage{graphicx}
\usepackage{braket}
\usepackage{amsthm}
\usepackage{comment}
\usepackage{array}
\usepackage{mathrsfs}

\usepackage[colorlinks=true,urlcolor=blue,citecolor=blue,linkcolor=blue]{hyperref}
\usepackage{physics}
\DeclareGraphicsExtensions{.eps,.ps,.pdf}
\newcommand{\mydash}{\mbox{---}}

\def \Tr {\text{Tr}}
\newcommand{\ie}{\hbox{\textit{i.e.}{}}}

\newcommand{\Hs}{\mathcal{H}}
\newcommand{\HSs}{\mathcal{B}}
\newcommand{\BHs}{\mathscr{B}}

\newcommand{\bcdot}{\boldsymbol{\cdot}}
\renewcommand{\vec}{\mathbf}

\newcommand{\group}{\mathcal{G}}
\newcommand{\pointg}[1]{\mathrm{#1}}

\newcommand{\Is}{\mathcal{I}_S}

\makeatletter
\g@addto@macro\bfseries{\boldmath}
\makeatother

\begin{document} 

\title{Dynamical decoupling of interacting spins through group factorization} 
\author{Colin Read}
\email{cread@uliege.be}
\affiliation{Institut de Physique Nucléaire, Atomique et de Spectroscopie, CESAM, University of Liège
\\
B-4000 Liège, Belgium}
\author{Eduardo Serrano-Ens\'astiga}
\email{ed.ensastiga@uliege.be}
\affiliation{Institut de Physique Nucléaire, Atomique et de Spectroscopie, CESAM, University of Liège
\\
B-4000 Liège, Belgium}
\author{John Martin}
\email{jmartin@uliege.be}
\affiliation{Institut de Physique Nucléaire, Atomique et de Spectroscopie, CESAM, University of Liège
\\
B-4000 Liège, Belgium}

\begin{abstract}
Dynamical decoupling (DD) is a well-known open-loop protocol for suppressing unwanted interactions in a quantum system, thereby drastically extending the coherence time of useful quantum states. In the original framework of evolution symmetrization, a DD sequence was shown to enforce a symmetry on the unwanted Hamiltonian, thereby suppressing it if the symmetry was inaccessible. In this work, we show how symmetries already present in the undesired Hamiltonian can be harnessed to reduce the complexity of decoupling sequences and to construct nested protocols that correct dominant errors at shorter timescales, using the factorization of DD symmetry groups into a product of its subgroups. We provide many relevant examples in various spin systems, using the Majorana constellation and point-group factorization to identify and exploit symmetries in the interaction Hamiltonian. Our framework recovers tailored pulse sequences developed in the context of NMR, including the classical Lee-Goldburg protocol, and further produces novel short and robust sequences capable of suppressing on-site disorder, dipole-dipole interactions, and more exotic many-body interactions in spin ensembles.   
\end{abstract}

\maketitle 
\section{Introduction}

Quantum hardware platforms often face limitations imposed by the short coherence times of their noisy constituents, which arise from unwanted interactions either among the constituents themselves or with their surrounding environment. These interactions result in decoherence and energy dissipation, which ultimately degrades the quantum state that encodes valuable information. Protecting quantum systems from these detrimental dynamics is essential for advancing various areas of quantum technology, including quantum computation~\cite{Preskill_2018quantum,NISQAlgo} and quantum metrology~\cite{He_2023,Shaji_2007,MetrologyDephasing_2018}.

\par A prominent tool for mitigating these effects in many experimental systems is dynamical decoupling (DD)~\cite{Viola_1998,Viola_1999,Viola_2013}, a specific kind of Hamiltonian engineering technique, closely related to NMR decoupling~\cite{LeeGoldburg_1965,Waugh_1968,Haeberlen_1968,Mehring_1972,REV8,Mansfield_1973,Burum_1979,Cory_1990,Levitt_1993,Levitt_2007,Mote_2016}, in which a carefully tailored sequence of pulses is applied to the system to periodically suppress unwanted interactions between constituents of the system and with the external environment. Dynamical decoupling was first introduced with the concept of symmetrization of a Hamiltonian~\cite{Viola_1999,Zanardi_1999}, since the first DD sequences were shown to project the unwanted Hamiltonian onto a specific symmetry sector of the space of operators acting on the system~; a symmetry group associated with this symmetrization process is called a \emph{decoupling group}. Although this approach has enabled the construction of highly robust sequences~\cite{Viola_2003, Tripathi_2024,Viola_2005} and quantum gates protected from finite-duration errors~\cite{DCG_2009_PRA,DCG_2009_PRL,DCG_2010_PRL}, the search for more efficient high-order DD sequences now mainly involves numerical methods~\cite{Lukin_2017,Lukin_2020,Cappellaro_2022,Lukin_2024,Lidar_2013,Biercuk_2009LODD} where the robustness requirement comes up as a constraint during optimization~\cite{Lidar_2013,Cappellaro_2022} or as conditions to be satisfied during a numerical search~\cite{Lukin_2020}. The pulse sequences obtained are generally no longer associated with a symmetry group.

\par In this work, we consider the original framework of evolution symmetrization~\cite{Zanardi_1999,Viola_1999} and show how factorization of a decoupling group as a product of its subgroups can be used to harness the symmetries of a Hamiltonian to reduce the complexity of a decoupling sequence. Moreover, we show that the same framework can be used to construct nested sequences that correct dominant errors at shorter timescales. We apply this idea to various spin systems, where the relevant decoupling groups are rotational symmetry groups (\emph{point groups}) and where the symmetries of a Hamiltonian are identified in its Majorana constellations~\cite{Serrano_2020,Read2025platonicdynamical}, in order to design novel short and robust group-based DD sequences. We show that the well-known Lee-Goldburg decoupling sequence~\cite{LeeGoldburg_1965}, which suppresses dipole-dipole interactions between pairs of spins~\cite{Mote_2016,Haeberlen_1968,Levitt_1993}, naturally emerges as a simple example of our framework. The new sequences derived in this work include a four-pulse (or eight-pulse Eulerian) sequence that suppresses on-site disorder and dipole-dipole interactions, with potential application in NMR spectroscopy~\cite{Waugh_1968,Mansfield_1973,Burum_1979,Cory_1990,Cappellaro_2022} and for coherence protection in interacting spin ensembles~\cite{Lukin_2020,Lukin_2024,Zhou_2020_metrology} and spin qudits~\cite{Omanakuttan_2024,Gross_2021,Gross_2024,OmanakuttanAndGross_2023}. These protocols are shown to outperform state-of-the-art sequences in relevant regimes of parameters for the decoupling of a disordered interacting spin ensemble. We also construct several short decoupling sequences capable of suppressing $K$-body interactions (with $K\leq 5$) in spin systems where the rotating-wave approximation is valid, as well as dephasing in "small" dimension qudits ($d\leq 6$). We believe that this intuitive framework opens a promising avenue for the design of pulse sequences that selectively decouple specific Hamiltonians, guided by the symmetries of the corresponding Majorana constellations~\cite{Levitt_2007,Mote_2016}.

\section{Basic concepts}
In this Section, we review the necessary notions of dynamical decoupling. We start with a reminder of the dynamical decoupling framework as introduced in Ref.~\cite{Viola_1999}. In Sec.~\ref{Sec.inaccsym}, we explain how decoupling sequences can be constructed by looking for \textit{inaccessible symmetries}. Finally, Sec.~\ref{secSU2} serves as a condensed pedagogical summary of our previous work and presents the knowledge necessary to fully understand the geometric concepts underlying each example presented in Sec.~\ref{Sec.Appli}. We show how applying the Majorana representation to operators provides an elegant geometric picture that can be used to uncover the inaccessible $\mathrm{SU}(2)$ symmetries of a given Hamiltonian. A more technical introduction to the Majorana representation of operators can be found in Ref.~\cite{Serrano_2020}, and the rigorous framework for the design of dynamical decoupling sequences consisting solely of global rotations using the Majorana constellations is presented in Ref.~\cite{Read2025platonicdynamical}.

\subsection{A reminder of dynamical decoupling}\label{Sec.Intro.DD}
We consider a quantum system $S$ coupled to an environment or bath $B$. The system-bath Hamiltonian can be written in Schmidt decomposition form,
\begin{equation}
H_{SB} = \sum_{\alpha} S_{\alpha} \otimes B_{\alpha},\label{eq:Schmidt_Hsb}
\end{equation}
where $S_{\alpha}$ and $B_{\alpha}$ are operators acting on the Hilbert spaces of the system and bath, respectively. From this decomposition, we define the \emph{interaction subspace} as the vector space spanned by the system operators $S_{\alpha}$, $\Is \equiv \mathrm{span}\qty(\qty{S_{\alpha}}) \subseteq \HSs\qty(\Hs_S)$. This subspace contains all system operators that couple to the environment, but it may also include terms acting solely on the system. Specifically, $H_{SB}$ can contain terms of the form $S_\alpha \otimes \mathds{1}_B$ (with $B_\alpha = \mathds{1}_B$). In the case where $S$ consists of several subsystems, for example spins in an ensemble, such terms can also describe unwanted interactions between the subsystems themselves.

\par The Hamiltonian~\eqref{eq:Schmidt_Hsb} entangles the system with the bath, resulting in the scrambling of quantum information initially stored in $S$. To reduce the resulting loss of coherence in $S$, we can apply a dynamical decoupling sequence of the form $\mydash P_1\mydash P_2\dots \mydash P_N$ to the system. This sequence consists of $N$ idealized pulses that are infinitely short and strong, where each $P_k$ represents the unitary operator corresponding to the $k$th pulse. The dashes ($\mydash$) denote free evolution intervals of duration $\tau_0$ between successive pulses. In the interaction picture with respect to the Hamiltonian implementing the DD pulses (toggling frame), the interaction Hamiltonian undergoes a series of unitary transformations
\begin{equation}
    H_{SB} \xrightarrow[\mydash P_1]{} P_1^{\dagger} H_{SB} P_1 \xrightarrow[\mydash P_2]{} P_1^{\dagger}  P_2^{\dagger} H_{SB} P_2 P_1 \xrightarrow[\mydash P_3]{}\cdots .
\end{equation}
An average Hamiltonian $H^{\text{av}}_{SB}$ can be calculated by performing a Magnus expansion in the toggling frame~\cite{Viola_1999}, and yields a series~\cite{Blanes_2009} $H^{\text{av}}_{SB} = \sum_{n=1}^{\infty}H_{SB}^{[n]}$. If decoherence is small enough ($T\norm{H_{SB}} \ll 1$ where $T \equiv N\tau_0$), then the series converges and can be approximated by the first-order term, given by 
\begin{equation}
     H_{SB}^{[1]} = \frac{1}{N}\sum_{k=1}^N (g_k^{\dagger} \otimes \mathds{1}_B) H_{SB} (g_k \otimes \mathds{1}_B)\label{eq.first.order}
\end{equation}
where we defined the propagator acting on $S$ at each step $k$ ($1\leq k\leq N$) of the sequence as 
\begin{equation}
    g_1\equiv \mathds{1}_S, \quad g_k \equiv 
    P_{k-1} P_{k-2} \ldots P_1.\label{eq:gk_propag}
\end{equation}
\par We now turn to the case where the set of operators $\group = \qty{g_k}_{k=1}^N$ forms a group of unitary transformations. Under this condition, the DD sequence effectively performs a \emph{symmetrization operation} $\Pi_{\group}$~\cite{Zanardi_1999,Viola_1999} on each system operator $S_{\alpha}$, 
\begin{equation}
    \Pi_{\group}: \HSs(\Hs_S) \to \HSs(\Hs_S): S \mapsto \Pi_{\group}(S) = \frac{1}{N}\sum_{k=1}^{N} g_k^{\dagger} S g_k .
    \label{eq:symmetrization}
\end{equation}
The trace-preserving map $\Pi_{\group}$ is a projector on a $\group$-invariant subspace of $\HSs(\Hs_S)$, the space of linear operators. Indeed, any element of $\group$ commutes with $\Pi_{\group}(S)$ for any $S\in \HSs(\Hs_S)$.  By linearity of \eqref{eq:symmetrization}, the image under $\Pi_{\group}$ of $\mathcal{I}_S$, $\Pi_{\group}(\mathcal{I}_S)$, is a vector subspace consisting of operators invariant under $\group$. If the group $\group$ is chosen such that the symmetrization map $\Pi_{\group}(\mathcal{I}_S)$ reduces the interaction subspace $\mathcal{I}_S$ to $\mathrm{span}(\qty{\mathds{1}_S})$, then $\group$ is referred to as a \emph{decoupling group} for $\mathcal{I}_S$. In this case, the first-order effective interaction Hamiltonian takes the form
\begin{equation}
    H_{SB}^{[1]} \propto \mathds{1}_S\otimes B ,
\end{equation}
where $B$ is an operator that acts only on the bath Hilbert space. In summary, if $\group$ is a decoupling group for $\mathcal{I}_S$, then the symmetrization map~\eqref{eq:symmetrization} eliminates all components of each system operator $S_\alpha$ in the interaction Hamiltonian~\eqref{eq:Schmidt_Hsb} that are not invariant under $\group$ (those with $\mathrm{Tr}[S_\alpha]=0$ going to $0$, the others to $\mathds{1}_S$). The only surviving contribution is proportional to the identity operator $\mathds{1}_S$, so the first-order effective interaction Hamiltonian reduces to a tensor product of the identity on the system with an operator acting only on the bath.

\par From a decoupling group $\group$, a sequence of pulses can be found that implements the operation \eqref{eq:symmetrization} by constructing a cyclic path on the Cayley graph $\mathrm{C}\qty(\group, \Gamma)$ of the group $\group$ with respect to some generating set $\Gamma= \qty{\gamma_{\lambda}}_{\lambda}$~\cite{Viola_2003,Viola_2013}~; in these graphs, each vertex corresponds to an element of the group and each edge to an element of $\Gamma$, which includes the different pulses used in the sequence, \ie, $\forall k$, $P_k= \gamma_{\lambda}$ for some $\lambda$. A cyclic path corresponds to a DD sequence in that each edge visited by the path corresponds to a pulse of the sequence and its location in the path corresponds to its location in the sequence. The only constraint imposed on the cyclic path is that it must visit each vertex the same number of times, so that many different DD sequences can be constructed that way. In particular, Eulerian paths, \ie, the ones that visit each edge exactly once, always exist~\cite{Bollobás1998} and it has been found that such paths correspond to particularly robust DD sequences~\cite{Viola_2003}.

\subsection{Inaccessible symmetry and decoupling group}\label{Sec.inaccsym}
Consider a potential decoupling group $\group$ for a quantum system with interaction subspace $\mathcal{I}_S$. To verify whether $\group$ decouples the system-bath interaction, one approach is to find a vector space $V$ containing $\mathcal{I}_S$ that is closed under unitary transformation by the elements of $\group$. Since $V$ is closed under unitary transformations of $\group$, it holds that $\Pi_{\group}(\mathcal{I}_S)\subseteq V$. Moreover, because $\Pi_{\group}(\mathcal{I}_S)$ consists solely of $\group$-invariant elements of $V$, it belongs to its $\group$-invariant subspace. When there is a group $\group$ for which the $\group$-invariant subspace of $V$ consists only of operators proportional to the identity, \emph{i.e.}, $\Pi_{\group}(V) = \mathrm{span} (\qty{\mathds{1}_S})$, we have, for any system operator $S$ belonging to the interaction subspace $\mathcal{I}_S \subset V$, that $\Pi_{\group}(S) \propto \mathds{1}_S$. This implies that $\group$ acts as a decoupling group for $\mathcal{I}_S$. When this condition is satisfied, we say that $\group$ is an \emph{inaccessible symmetry} of $V$. In conclusion, an inaccessible symmetry of a vector space containing the interaction subspace is associated with a decoupling group.

\subsection{\texorpdfstring{$\mathrm{SU(2)}$}{Lg} symmetries}
\label{secSU2}
\par For an ensemble of interacting spins, the most common dynamical decoupling protocols are composed of global $\mathrm{SU(2)}$ transformations, where each spin undergoes the same unitary evolution $U = e^{-i\theta \hat{n}\bcdot\mathbf{J}}$. Here $\mathbf{J}=(J_x,J_y,J_z)$ are the angular momentum operators of a single spin $j$. The action of any $U \in \mathrm{SU}(2)$  over linear operators by conjugation can be thought as a physical rotation $\mathrm{R}(\hat{n},\theta)\in\mathrm{SO(3)}$, where $\mathrm{R}(\hat{n},\theta)$ represents a rotation of angle $\theta$ around a certain axis $\hat{n}$\footnote{$\mathrm{SU}(2)$ is the double cover of $\mathrm{SO}(3)$, where both transformations $\pm U = \pm e^{-i\theta \hat{n}\bcdot\mathbf{J}}$ can be associated to the same rotation $\mathrm{R}(\hat{n},\theta)$. Note that the action of $\pm U$ by conjugation on an operator $A$ gives the same result because $(\pm U) A (\pm U)^{\dagger} = U A U^{\dagger}$. }.

\par For a given interaction subspace $\Is$, we now consider a vector space $V$ closed under $\mathrm{SU}(2)$ transformations such that $\mathcal{I}_S\subseteq V$. Then, $V$ can be written in terms of irreducible representations [vector subspaces closed under $\mathrm{SU}(2)$ transformations] as~\cite{Varshalovich_1988}
\begin{equation}
    V = \bigoplus_{(L,\alpha)} \BHs^{(L,\alpha)} .\label{eq.decomp.irrep}
\end{equation}
Each subspace $\BHs^{(L,\alpha)}$ of dimension $2L+1$ is called an $L$-\emph{irrep}. The extra index $\alpha$ may be used to distinguish between different irreps of the same dimension.

\par As shown in Ref.~\cite{Serrano_2020}, any Hermitian operator in the subspace $\BHs^{(L,\alpha)}$ can be represented geometrically as a two-color constellation of $2L$ points on the sphere\footnote{If the operator is not Hermitian, two distinct Majorana constellations are needed, one representing the Hermitian part and the other the anti-Hermitian component. See Ref.~\cite{Read2025platonicdynamical} for more details.}, which we call its \emph{Majorana  constellation}~\cite{Serrano_2020,Majorana_1932}. The constellation consists of exactly $L$ pairs of antipodal points (referred to as \emph{stars}) on a sphere of a fixed radius, with each pair comprising one black and one red star. The specific arrangement of these bicolored pairs, along with the radius,  uniquely determines the corresponding operator. A general operator in $V$ can be represented by a collection of such constellations, one for each irreducible component $\BHs^{(L,\alpha)}$ in the decomposition~\eqref{eq.decomp.irrep}.

\begin{figure}[t!]
    \includegraphics[width=0.85\linewidth]{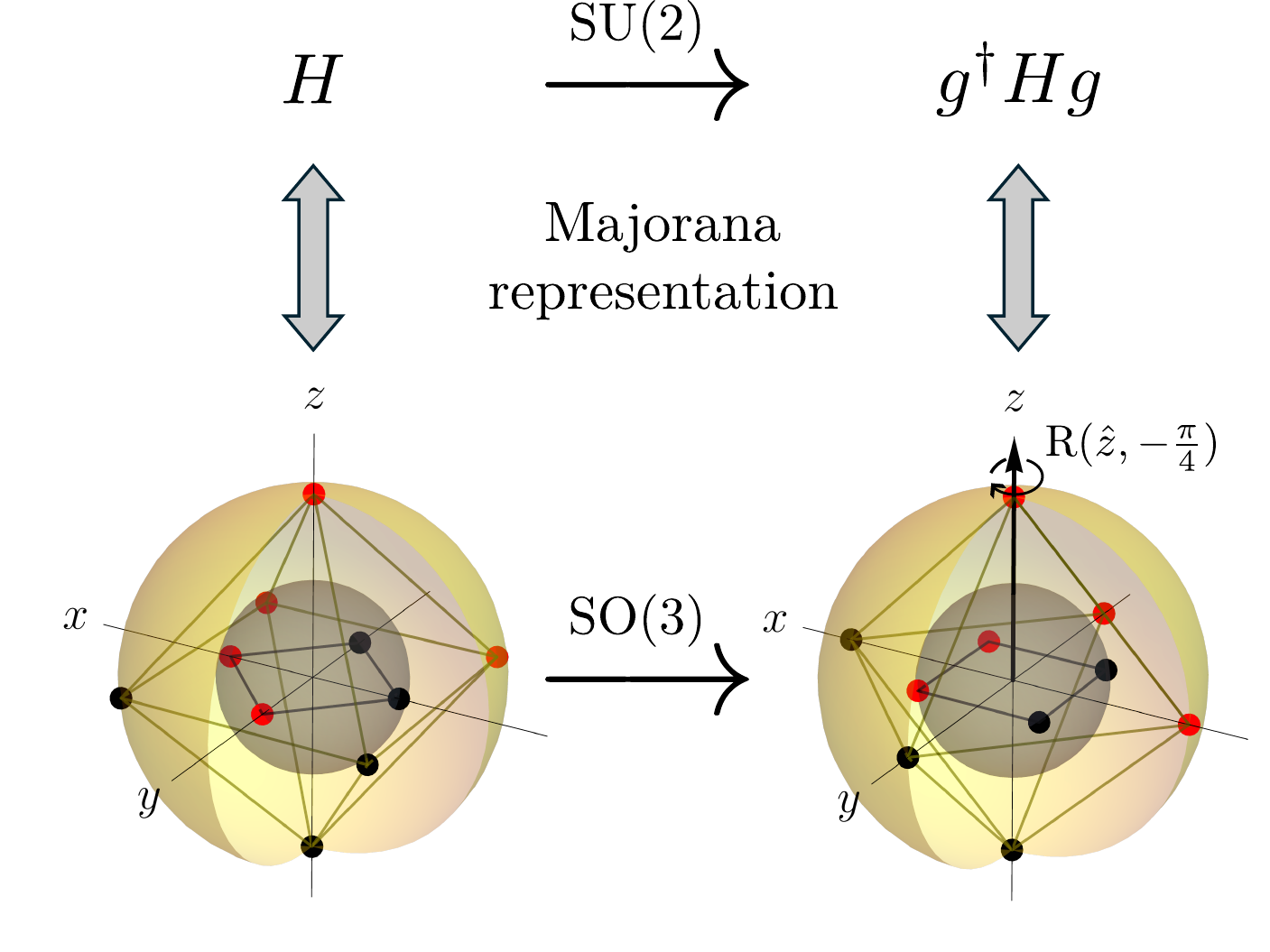}
    \caption{Majorana constellation of an operator $H$. The diagram also illustrates the effect of an $\mathrm{SU(2)}$ transformation $g^\dagger$ on the operator and its corresponding $\mathrm{SO(3)}$ rotation $\mathrm{R}(\hat{n},\theta)$ acting on the constellations associated with $H$. In this illustrative example, the constellation is rotated by an angle $\theta=-\pi/4$ about the $z$ axis. Figure reproduced from Ref.~\cite{Read2025platonicdynamical}.}
    \label{fig1:Constellation}
\end{figure}

\par One of the main appeals of the Majorana representation lies in the fact that applying a global $\mathrm{SU(2)}$ transformation on an operator $H\in V$ is equivalent to applying the corresponding $\mathrm{SO(3)}$ rotation on each Majorana constellation of $H$ (see Fig.~\ref{fig1:Constellation}). Consequently, the symmetries of an operator under $\mathrm{SU}(2)$ transformations can be studied by investigating the rotational symmetries of the corresponding set of constellations. More precisely, an operator $S$ is invariant under a $\mathrm{SU}(2)$ unitary transformation $g$ (\emph{i.e.}, $g^\dagger S g = S$) if the corresponding $\mathrm{SO}(3)$ rotation leaves each Majorana constellation unchanged, except possibly for an even number of color swaps between antipodal star pairs. This is because for any operator $S \in \BHs^{(L)}$, swapping the colors of a pair of antipodal stars changes the sign of the operator, such that swapping the colors of $N$ pairs of stars adds a factor $(-1)^N$ to the operator~\cite{Serrano_2020,Read2025platonicdynamical}.

\par Because the accessible rotational symmetries of a given constellation are limited by its number of points, it is possible to identify inaccessible $\mathrm{SU(2)}$ symmetries of operators using this representation. For more details on the Majorana representation of operators and its use to detect inaccessible \emph{point group} symmetries, and thus DD sequences for different physical systems, see~\cite{Read2025platonicdynamical}. For the first few irreps $\BHs^{(L)}$ (for $L\leq 7$), we specify in Table~\ref{tab1:Point.Groups} the sets of largest accessible symmetries $\mathcal{F}_{\mathrm{max}}(\BHs^{(L)})$, meaning that any rotational symmetry contained as a subgroup of at least one element of this set is accessible to the irrep. We also specify several smallest inaccessible symmetries. Note that if a point group $\group$ is an inaccessible symmetry, then any point group containing $\group$ as a subgroup is also inaccessible.

\begin{table*}[t]
    \centering
    \setlength{\tabcolsep}{8pt}
    \renewcommand*{\arraystretch}{1.5}
    \begin{tabular}{c||c|c|c|c|c|c|c|c}
        $L$ & 0 & 1 & 2 & 3 &4 & 5 & 6 & 7\\\hline
        $\mathcal{F}_{\mathrm{max}}(\BHs^{(L)})$ & $\mathrm{SO(3)}$ & $\pointg{C}_{\infty}$ & $ \pointg{D}_{\infty} $ & $\left\{ \pointg{C}_{\infty}, \pointg{D}_3 ,  \pointg{T} \right\}$ & $\qty{\pointg{D}_{\infty} , \pointg{O}}$ & $\qty{\pointg{C}_{\infty},\pointg{D}_{5}}$ & $\left\{ \pointg{D}_{\infty}, \pointg{O} ,  \pointg{I} \right\}$& $\qty{\pointg{C}_{\infty},\pointg{D}_{7},\pointg{T}}$\\
        Smallest inaccessible $\group$ &  none & $\pointg{D}_{2}$ & $\qty{\pointg{T}, \pointg{I}}$ & $\qty{\pointg{D}_4,\pointg{O}, \pointg{I}}$ & $\pointg{I}$ & $\qty{\pointg{D}_6,\pointg{T}, \pointg{I}}$ & none & $\qty{\pointg{D}_8,\pointg{O}, \pointg{I}}$
    \end{tabular}
    \caption{First row: Sets of largest point groups for irreps of $\mathrm{SU}(2)$, $\mathcal{F}_{\mathrm{max}}(\BHs^{(L)})$, of different dimensions $2L+1$. Second row: Set of the smallest symmetries inaccessible to these irreps. Any group that contains one of these groups as a subgroup is itself an inaccessible symmetry. A similar analysis can be done for higher $L$ by searching for the accessible and inaccessible point groups.}
    \label{tab1:Point.Groups}
\end{table*}

\par To illustrate our approach, let us consider the simple case of a disordered and interacting spin ensemble, where the unwanted Hamiltonian is given by $H = H_{\mathrm{dis}} + H_{\mathrm{dd}}$, with 
\begin{equation}
\begin{aligned}
  H_{\mathrm{dis}} ={}& \sum_i\delta_i\hat{m}_i\bcdot\mathbf{S}^i , \\
  H_{\mathrm{dd}} ={} &
\sum_{ij}\Delta_{ij}\qty[3\qty(\hat{e}_{ij}\bcdot\mathbf{S}^i)\qty(\hat{e}_{ij}\bcdot\mathbf{S}^j) - \mathbf{S}^i\bcdot\mathbf{S}^j].
\label{eq:Hamiltonian} 
\end{aligned}
\end{equation}
where $\vec{S}^i = (S^i_x,S^i_y,S^i_z)$ is the angular momentum vector of the $i$th spin. In~\eqref{eq:Hamiltonian}, $H_{\mathrm{dis}}$ describes disorder in the ensemble (\ie\, each spin rotating at different frequencies $\delta_i$ around different axes $\hat{m}_i$) while $H_{\mathrm{dd}}$ describes dipole-dipole interactions between pairs of spins ($i$,$j$) with relative orientation $\hat{e}_{ij}$ and coupling factor $\Delta_{ij}$. In the interaction-dominated regime ($\norm{H_{\mathrm{dd}}}\gg \norm{H_{\mathrm{dis}}}$), this model describes the main sources of decoherence in nuclear spin ensemble and has been widely used to construct decoupling sequences for high-resolution NMR spectroscopy~\cite{Waugh_1968,REV8,Mansfield_1973,Burum_1979,Cory_1990} and Hamiltonian engineering~\cite{Cappellaro_2022}. In the disorder-dominated regime ($\norm{H_{\mathrm{dis}}}\gg \norm{H_{\mathrm{dd}}}$), the model is relevant for electronic spin ensembles~\cite{Tyryshkin_2003,Choi_2017,Kucsko_2018,Lukin_2020}. 
\par
\begin{figure}[t]
    \centering
    \includegraphics[width=0.9\linewidth]{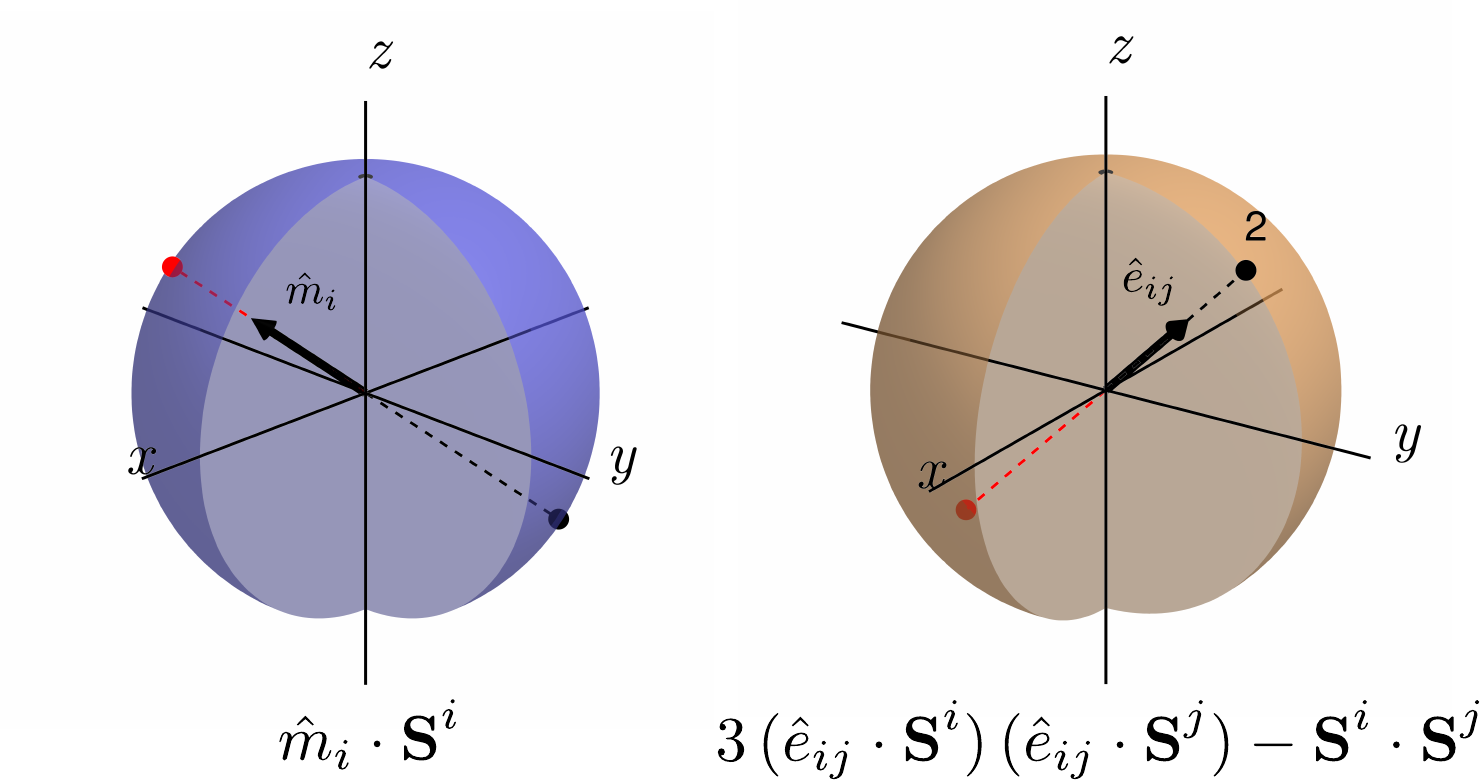}
    \caption{Majorana constellations of Hamiltonians $H\propto \hat{m}_i\bcdot \vec{S}^i$ (left) and $H\propto 3(\hat{e}_{ij}\bcdot \vec{S}^i)(\hat{e}_{ij}\bcdot \vec{S}^j) - \vec{S}^i\bcdot\vec{S}^j$ (right). The label $2$ next to the black star in the right-hand constellation indicates a degeneracy where two pairs of antipodal stars occupy the same position on the sphere.}
    \label{fig2:const.H}
\end{figure}

 Each term in the sum defining $H_{\mathrm{dis}}$ (resp.\ $H_{\mathrm{dd}}$) belongs to a $L=1$-irrep (resp.\ $L=2$-irrep), such that its associated set of constellations contains exactly one constellation with one pair (resp.\ two pairs) of antipodal stars (see Fig.~\ref{fig2:const.H}). Referring to Table~\ref{tab1:Point.Groups}, we observe that the tetrahedral point group $\mathrm{T}$ is inaccessible to the $L=2$ irrep. Because $\mathrm{T}$ contains $\mathrm{D}_2$ as a subgroup, and $\mathrm{D}_2$ is inaccessible to the $L=1$ irrep, it follows that $\mathrm{T}$ is also inaccessible to the $L=1$ irrep. This allows us to construct an Eulerian DD sequence on the Cayley graph of the tetrahedral group, resulting in the sequence known as $\mathrm{TEDD}$,  first introduced in Ref.~\cite{Read2025platonicdynamical}. A similar reasoning can be used for more exotic Hamiltonians whose decomposition involves $L\geq 3$-irreps, and one finds that relevant DD sequences can be constructed from the Cayley graphs of the different Platonic symmetry groups; they are thus referred to as Platonic dynamical decoupling sequences and denoted by $\mathrm{xEDD}$, where $\mathrm{x}\in\qty{\mathrm{T},\mathrm{O},\mathrm{I}}$ indicates the symmetry group.

\section{Multisymmetrization}\label{Sec.MS}
\subsection{Nested sequences}
\begin{figure}[t]
    \centering
    \includegraphics[width=0.9\linewidth]{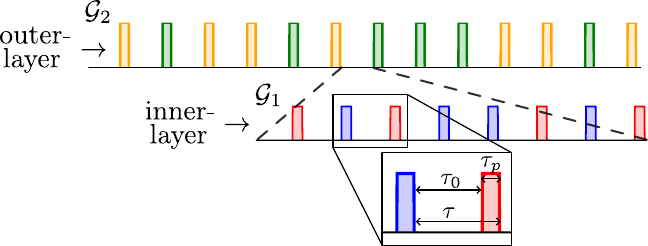}
    \caption{Visual representation of the nesting process for multisymmetrization. Each stick represents a pulse ($P_k$) and the different colours represent the different generators ($\gamma_\lambda$). The waiting time between successive pulses is set to $\tau_0$ and the duration of each pulse to $\tau_p$, so that the total pulse interval is $\tau = \tau_0 + \tau_p$. }
    \label{fig3:nesting}
\end{figure}

Consider now two groups of unitary operators, $\group_1$ and $\group_2$, and their corresponding sequence of pulses that implement their respective symmetrization operations~\eqref{eq:symmetrization}. A new pulse sequence can be constructed by \emph{nesting} one sequence into the other~\cite{Khodjasteh_2005,Tripathi_2024}. This is done by taking one of the pulse sequences (called the \emph{outer-layer} sequence) and replacing the waiting time between the pulses by an iteration of the second pulse sequence (the \emph{inner-layer} sequence). We show a graphical representation of the nesting process in Fig.~\ref{fig3:nesting}, where the sequence corresponding to the group $\mathcal{G}_2$ is the outer-layer sequence, and the sequence corresponding to $\mathcal{G}_1$ is the inner-layer sequence. By applying this sequence, any system operator $S$ is first symmetrized by the inner-layer group, and then symmetrized again by the outer-layer group, as follows
\begin{equation}
S \mapsto S' = \Pi_{\group_1}(S), \quad S' \mapsto S'' = \Pi_{\group_2}(S').
\end{equation}
In total, the operator $S$ undergoes two successive (potentially non-commuting) quantum operations. 

In some cases, the two groups form a \emph{factorization} of another group, $\group_2\group_1 = \group'$\footnote{The (Frobenius) product of two groups $\group_1=\qty{g^1_i}_{i=1}^{\abs{\group_1}}$ and $\group_2=\qty{g^2_i}_{i=1}^{\abs{\group_2}}$ is defined as the set of elements obtained by multiplying all pairs of group elements together, $\group_2\group_1 = \qty{g | g=g_i^2g_j^1 \text{ with } g_i^1 \in \group_1 \text{ and } g_j^2 \in \group_2}$~\cite{ballester2010products}.}. When the groups are \emph{complementary} to each other (\emph{i.e.}, $\group_1$ and $\group_2$ only intersect in the identity and therefore $|\group_1 \cap \group_2|=1$), the product formula~\cite{ballester2010products} states that $|\group'|= |\group_1||\group_2| / |\group_1 \cap \group_2|= |\group_1||\group_2| $. This implies that each element $g\in \group'$ has a unique decomposition $g=g^2 g^1$ with $g^a \in \group_a$ for $a=1,2$. Due to these properties, successive symmetrization operations of an operator $S$ with respect to these two groups are reduced to a single effective symmetrization operation with respect to $\group'$,
\begin{equation}
\label{Eq.nested}
\begin{aligned}    
\Pi_{\group_2}\qty[\Pi_{\group_1}\qty(S)] ={} &\frac{1}{\abs{\group_1}\abs{\group_2}} \sum_{i=1}^{|\group_1|} \sum_{j=1}^{|\group_2|} g_j^2 g_i^1 S (g_j^2 g_i^1)^{-1}
\\
={} & 
\frac{1}{\abs{\group'}} \sum_{i=1}^{|\group'|} g_i' S (g_i^{\prime})^{-1}
\\
={} & 
\Pi_{\group'}\qty(S) .
\end{aligned}
\end{equation}
The operator $S$ is thus projected onto a $\group'$-invariant subspace through what we refer to as a \emph{multisymmetrization} operation. Although this result is general for any abstract group, we focus on point groups in this work.

\par In the case where $\group_1\cap\group_2$ is not trivial, then again, by the product formula and because $\group'$ is a group, each element of $\group'$ will appear $\abs{\group_1 \cap\group_2}$ times in the symmetrization process~\cite{Rotman_2012}. Thus, we have
\begin{equation}
    \Pi_{\group_2}\qty[\Pi_{\group_1}\qty(S)] = \frac{\abs{\group_1\cap \group_2}}{\abs{\group_1}\abs{\group_2}}\sum_{i=1}^{|\group'|} g_i' S (g_i^{\prime})^{-1} = \Pi_{\group'}\qty(S) ,
\end{equation}
which does not compromise the multisymmetrization and leaves more freedom in the choice of subgroups used in the nesting process. However, the choice of two subgroups that only intersect in the identity minimizes the number of pulses used in the sequence, which is generally favorable. For this reason, we focus on factorizations where the subgroups are complementary. These are listed in Table~\ref{tab2:point_group_factorizations}  for exceptional point groups. To search exhaustively for such factorizations, we examine the subgroup structure of a group~\cite{conway1985finite}. We describe the method of identifying factorizations with complementary subgroups in Appendix~\ref{Ap.Subgroups}. The same method can be applied to more general groups. Finally, note that the product of $\group_1$ and $\group_2$ is itself a group if and only if $\group_1 \group_2 = \group_2\group_1$~\cite{walker_1987,Ledermann_1976}, which means that the inner and outer layers can be safely interchanged without compromising multisymmetrization.

Two possible applications of this operation are described in the next section.

\subsection{Applications of sequence nesting}

\subsubsection{Reducing a decoupling group using existing symmetries}\label{application1}
Consider a quantum system with a Hilbert space $\Hs_S$, and an interaction subspace $\mathcal{I}_S \subset \HSs\qty(\Hs_S)$ that is invariant under a nontrivial group of unitary operators $\group_{1}$. Now suppose that we apply a symmetrization operation associated with a second group $\group_2$ to the system. Since $\mathcal{I}_S$ is already invariant under $\group_1$, any operator $S\in\mathcal{I}_S$ will transform as
\begin{equation}
    S \mapsto \Pi_{\group_2}(S) =  \Pi_{\group_2}\qty[\Pi_{\group_1}\qty(S)].
\end{equation}
As noted above, if the product $\group_1\group_2$ forms a group $\group'$, then this single symmetrization procedure effectively projects $S$ onto a $\group'$-invariant subspace of $\HSs\qty(\Hs_S)$. If $\group'$ is an inaccessible symmetry, then the sequence implementing the $\group_2$ symmetrization procedure will average the operator $S$ to zero or to a multiple of the identity (see Sec.~\ref{Sec.Intro.DD}). From this brief analysis, we conclude that it is possible to reduce the complexity of a DD sequence by exploiting the symmetries already present in the interaction subspace. If a decoupling group can be factorized into a product of two of its subgroups, one of which corresponds to a symmetry of the interaction subspace, then the remaining subgroup is a decoupling group. This idea of exploiting existing symmetries to reduce the size of a decoupling group was already presented in Ref.~\cite{Viola_2003} in the context of reducing the length of an Eulerian DD sequence, where it was noted that if the interaction subspace is $\group_0$-invariant, where $\group_0$ is a normal subgroup of a decoupling group $\group$, then the DD sequence can be constructed on the Cayley graph of the quotient group $\group/\group_0$, of order $\abs{\group}/\abs{\group_0}$. In fact, the CPMG sequence~\cite{Viola_1999,Viola_1998} for a dephasing qubit and the $d\mathrm{X}_d$ sequence family~\cite{Tripathi_2024} for dephasing qudits can be seen as two simple examples of this.

\subsubsection{Hierarchical decoupling}\label{application2}
Let the group $\group'$ serve as a decoupling group for all unwanted interactions that lead to decoherence. Because different interactions can induce decoherence on different timescales, those associated with faster dynamics should be suppressed with greater priority~\cite{Lukin_2024}. This can be achieved by carefully choosing the group factorization so that the subgroup $\group_1$ targets and suppresses the dominant, fast-acting interaction terms. In this way, we can implement the symmetrization with respect to $\group'$ while ensuring that the most critical decoherence channels are addressed in priority through the structure of the subgroup hierarchy.

\begin{table}[t!]
\centering
\setlength{\tabcolsep}{8pt}
\renewcommand*{\arraystretch}{1.5}
\begin{tabular}{l|c|l}
Group & Order & Factorization \\
\hline\hline
$\mathrm{T}$ (tetrahedral) & 12 & $ \mathrm{D}_2 \mathrm{C}_3$ \\
\hline
$\mathrm{O}$ (octahedral) & 24 & $\mathrm{T} \mathrm{C}_2$, $\mathrm{D}_4 \mathrm{C}_3$, $\mathrm{D}_3 \mathrm{D}_2$, $\mathrm{D}_3 \mathrm{C}_4$ \\
\hline
$\mathrm{I}$ (icosahedral) & 60 & $\mathrm{T} \mathrm{C}_5 $  \\
\end{tabular}
\caption{Complementary subgroup factorizations of the tetrahedral, octahedral, and icosahedral groups. Each factorization $\group = \group_1 \group_2$ is also valid in reverse order: $\group = \group_2 \group_1$.}
\label{tab2:point_group_factorizations}
\end{table}

\section{Finite-duration pulses}\label{sec:finite_duration}

\subsection{Generalized symmetrization operation}
When the approximation of the infinitely fast and strong pulses is not applicable, the derivation of the first-order average Hamiltonian presented in Sec.~\ref{Sec.Intro.DD} no longer holds, and the effect of the DD sequence on the Hamiltonian can no longer be reduced to the symmetrization operation~\eqref{eq:symmetrization}. 

\par In the most general case, where each pulse $P_k$ is smoothly implemented in the time interval $[(k-1)\tau,k\tau]$ by a unitary propagator $p_k(t)$, with $p_k(\tau) = P_k$, the first-order Hamiltonian can be calculated explicitly~\cite{Viola_2003,Viola_2013}. The pulse interval $\tau=\tau_p+\tau_0$ thus includes both the waiting time between successive pulses (denoted $\tau_0$ in Fig.~\ref{fig3:nesting}) and the duration of the pulse $\tau_p$. In this scenario, each system operator $S_{\alpha}$ undergoes the quantum operation $Q_{\mathcal{G}}\,:\HSs(\Hs_S)\to \HSs(\Hs_S)$ which acts as follows
\begin{equation}
   S \mapsto Q_{\mathcal{G}}(S) = \frac{1}{N}\sum_{k=1}^Ng_k^{\dagger}\qty[\frac{1}{\tau} \int_0^{\tau} p_{k}^{\dagger}(t)\,S\, p_k(t) dt]g_k.
\end{equation}
More specifically, consider a dynamical decoupling (DD) sequence that traces an Eulerian path on the Cayley graph of a group $\mathcal{G}$, constructed with respect to a generating set $\Gamma = \{ \gamma_{\lambda} \}_{\lambda=1}^{|\Gamma|}$. Each pulse $P_k$ in the sequence is an element of $\Gamma$ and is associated with a time-dependent propagator $p_k(t)$, which belongs to the set of propagators
\begin{equation}
   U \equiv \left\{ u_{\lambda}(t) : u_{\lambda}(\tau) = \gamma_{\lambda} \right\}_{\lambda=1}^{|\Gamma|},
\end{equation}
where each $u_\lambda(t) \in U$ represents a fixed physically realizable control operation that implements the generator $\gamma_\lambda$ over a time interval $\tau$. These propagators are dependent on experimentally accessible pulse profiles, including details such as their shape, amplitude, and duration. In this case, the map $Q_{\mathcal{G}}$ can be written as~\cite{Viola_2003,Viola_2013}
\begin{equation}
    S \mapsto Q_{\group}(S) = \Pi_{\group}\qty[F_{\Gamma}^{(\tau)}(S)]
\end{equation}
where $\Pi_{\group}$ is the symmetrization operation as defined in Eq.~\eqref{eq:symmetrization} and $F_{\Gamma}^{(\tau)}\,:\HSs(\Hs_S)\to \HSs(\Hs_S)$ acts as follows
\begin{equation}
    S \mapsto F_{\Gamma}^{(\tau)}(S) = \frac{1}{\abs{\Gamma}}\sum_{u\in U} \frac{1}{\tau} \int_0^{\tau} u^{\dagger}(t)\, S\, u(t)dt,
\end{equation}
where the sum runs over the set of pulses $U$ that implement the elements of the generating set $ \Gamma$. Equivalently, we can write 
 \begin{equation}
    Q_{\group}(S) = \frac{1}{\abs{\Gamma}}\sum_{\lambda=1}^{|\Gamma|} \Pi_{\group}\qty[F_{\lambda}^{(\Gamma,\tau)}(S)]\label{eq:Qg}
\end{equation}
where $ F_{\lambda}^{(\Gamma,\tau)}(S) \equiv F_{ \{ \gamma_{\lambda} \}  }^{(\tau)}(S)  $ denotes the implementation of $F$ over a singleton $\{ \gamma_{\lambda} \}\in \Gamma$. Physically, the operator $F_{\lambda}^{(\Gamma,\tau)}(S)$ represents the finite-duration error generated by the presence of the system operator $S$ during the generating pulse $u_{\lambda}(t)$. As can be understood from Eq.~\eqref{eq:Qg}, the finite-duration errors of each generator will be symmetrized independently~\eqref{eq:symmetrization}. Note that if $u_{\lambda}(t)$ has support on the time interval $[\tau-\tau_p,\tau]$ (where $\tau-\tau_p$ is the waiting time $\tau_0$), one can also write 
\begin{equation}
    F_{\lambda}^{(\Gamma,\tau)}(S) = \frac{\tau_0}{\tau}S + \frac{\tau_p}{\tau} f_{\lambda}^{(\Gamma,\tau_p)}(S)
\end{equation}
where 
\begin{equation}
    f_{\lambda}^{(\Gamma,\tau_p)}(S) = \frac{1}{\tau_p} \int_{\tau-\tau_p}^{\tau} u_{\lambda}^{\dagger}(t)\, S\, u_{\lambda}(t)dt.
\end{equation}

\par For some group $\mathcal{G}$, we can now define a correctable subspace $\mathcal{C}_{\group}$ spanned by all the system operators $S_\alpha$ satisfying $\Pi_{\group}(S_\alpha)\propto \mathds{1}_S$. In the ideal pulse regime, the decoupling group $\group$ should be chosen such that $\mathcal{I}_S \subseteq \mathcal{C}_{\group}$. In the finite-pulse regime, on the other hand, we must make sure that $F_{\lambda}^{(\Gamma,\tau)}(\mathcal{I}_S)$ also belongs to that correctable subspace for all generating pulses $u_{\lambda}(t)$. This requirement has several implications in the context of the two applications discussed in Subsec.~\ref{application1} and \ref{application2}, as described below.
\subsection{Applications}
\subsubsection{Leveraging existing symmetries}\label{Sec.fin.dur.lev}

Consider an interaction subspace $\mathcal{I}_S$ invariant under the action of the group $\group_1$ and contained in the correctable subspace $\mathcal{C}_{\group'}$ of a certain group $\group'$. As explained previously, in the case of ideal pulses, if there exists a group $\group_2$ that satisfies $\group'=\group_2\group_1$, then this is a decoupling group for $\mathcal{I}_S$, \ie, $\mathcal{I}_S\subseteq \mathcal{C}_{\group_2}$.

\par In the finite-pulse regime, when one chooses an Eulerian path on the Cayley graph $C(\group_2,\Gamma_2)$, every operator $S$ in $\mathcal{I}_S$ undergoes the operation~\eqref{eq:Qg} 
\begin{equation}
    S \mapsto Q_{\group_2}(S) = \frac{1}{\abs{\Gamma_2}}\sum_{\lambda}\Pi_{\group_2}[F^{(\Gamma_2,\tau)}_{\lambda}(S)]
\end{equation}
where $\Gamma_2 =\qty{\gamma_{2,\lambda}}_{\lambda}$ are the chosen generators and are implemented by the generating pulses $\qty{u_{2,\lambda}(t): u_{2,\lambda}(\tau)=\gamma_{2,\lambda}}_{\lambda}$ and $\tau$ is the pulse interval. Although $S = \Pi_{\group_1}(S)$ $\forall S\in\mathcal{I}_S$, the symmetrization $\Pi_{\group_1}(\cdot)$ generally does not commute with $F^{(\Gamma_2,\tau)}_{\lambda}(\cdot)$, which means that $F^{(\Gamma_2,\tau)}_{\lambda}(\mathcal{I}_S)$ is generally not $\group_1$ invariant. Furthermore, $F^{(\Gamma_2,\tau)}_{\lambda}(\mathcal{I}_S)$ might leak out of the correctable subspace of $\group'$. The robustness of these designs to finite-duration pulses must then be addressed on a case-by-case basis, and specific requirements on the shape and implementation of the generating pulses might be required to ensure robustness.

\subsubsection{Hierarchical decoupling}

Consider a $\group'=\group_2\group_1$-symmetrization procedure implemented through two nested layers of symmetrization: an inner layer with respect to the group $\group_1$ and an outer layer with respect to the group $\group_2$. The nested DD sequence is constructed using Eulerian sequences on the Cayley graphs $C(\group_1,\Gamma_1)$ and $C(\group_2,\Gamma_2)$, where each $\Gamma_j = \{ \gamma_{j,\lambda} \}_{\lambda}$ is a generating set for the group $\group_j$ and $U_j = \{ u_{j,\lambda}(t) \}_{\lambda}$ are the generating pulses. 

\par We first apply the Eulerian sequence associated with $\group_1$, with a total duration $T = \abs{\group_1}\abs{\Gamma_1}\tau_1$ where $\tau_1=\tau_0^{(1)}+\tau_p^{(1)}$ is the interval between two pulses that includes the duration of a pulse $\tau_p^{(1)}$ and the possible waiting time between two pulses $\tau_0^{(1)}$. This DD sequence will effectively implement the operation $S \mapsto S' = Q_{\group_1}(S)$ defined in Eq.~\eqref{eq:Qg} with 
\begin{equation}
    F^{(\Gamma_1,\tau_1)}_{\lambda}(S) = \frac{1}{\tau_1}\int_0^{\tau_1} u_{1,\lambda}^{\dagger}(t) S u_{1,\lambda}(t)  dt.
\end{equation}
This sequence is now nested into the Eulerian sequence associated with $\group_2$, with a total duration $T' = \abs{\group_1}\abs{\group_2}(T+\tau_p^{(2)})$, where each pulse interval of duration $T$ is replaced by the $\group_1$ Eulerian sequence [see Fig.~\ref{fig3:nesting}] and where $\tau_p^{(2)}$ is the pulse duration for the second sequence. 

In the ideal pulse regime ($\tau_p^{(2)}=0$), the outer-layer sequence effectively symmetrizes the operator $S'$, and the total effect of the nested sequence is to induce the quantum operation
\begin{equation}
    S \mapsto S'' = \Pi_{\group_2}(S') = \frac{1}{\abs{\Gamma_1}}\sum_{\lambda=1}^{\abs{\Gamma_1}}\Pi_{\group_2}\qty[\Pi_{\group_1}\qty[F^{(\Gamma_1,\tau_1)}_{\lambda}(S)]].
\end{equation}
Therefore, as long as the finite-duration errors of the pulses of the inner-layer sequence belong to the correctable subspace of $\group'=\group_2\group_1$, \ie, $F^{(\Gamma_1,\tau_1)}_{\lambda}(\mathcal{I}_S)\subseteq \mathcal{C}_{\group'}$, the nested sequence decouples the interaction subspace.

\par In the case where $\tau_p^{(2)}>0$, it must be taken into account that the error Hamiltonian during the implementation of the pulses of the outer-layer sequence is not symmetrized by the inner-layer, so that the nested sequence effectively implements~\cite{Khodjasteh_2005, Khodjasteh_2007}
\begin{multline}
    S \mapsto S'' = \frac{1}{\abs{\Gamma_2}}\sum_{\lambda=1}^{\Gamma_2}\Pi_{\group_2}\left[\frac{T}{T+\tau_p^{(2)}}S'\right. \\ + \left.\frac{\tau_p^{(2)}}{T+\tau_p^{(2)}}f^{(\Gamma_2,\tau_p^{(2)})}_{\lambda}(S)\right],
\end{multline}
where 
\begin{equation}
    f^{(\Gamma_2, \tau_p^{(2)})}_{\lambda}(S) = \frac{1}{\tau_p^{(2)}}\int_T^{T+\tau_p^{(2)}} u_{2,\lambda}^{\dagger}(t) \, S\, u_{2,\lambda}(t)dt
\end{equation}
and $S'= Q_{\group_1}(S)$. As a consequence, the finite-duration errors of the outer-layer pulses do not undergo the multisymmetrization, which reduces the robustness of our schemes. 

\section{Applications to spin systems and decoherence models}\label{Sec.Appli}
In this section, we apply the framework presented in Secs.~\ref{Sec.MS} and \ref{sec:finite_duration} to several spin systems, using the Majorana representation of operators to identify existing symmetries and factorisation of point groups into smaller subgroups to construct more efficient DD sequences. We first address the decoupling of dipole-dipole interactions in a spin ensemble, followed by the case including disorder. Next, we consider $K$-body multilinear interactions in a spin ensemble. Finally, we address qudit decoupling from a dephasing bosonic bath and explore hierarchical decoupling in spin qudits. A summary of the most relevant DD sequences that we develop in the following is given in Tables~\ref{Table.4.leverage} and~\ref{Table.5.concatenation} in the Appendix~\ref{App.C.summary}.

\subsection{Dipole-dipole interaction}

As a first example, consider the Hamiltonian $H_{\mathrm{dd}}$ in Eq.~\eqref{eq:Hamiltonian} under the Rotating Wave Approximation (RWA), where we assume that all dipole-dipole interactions are aligned along the same direction, say $\hat{e}_{ij} =\hat{z}$. This Hamiltonian reads
\begin{equation}
    H_{\mathrm{dd}}^{\mathrm{RWA}} = \sum_{i<j}\Delta_{ij} \qty(3S_z^iS_z^j - \mathbf{S}^i \bcdot\mathbf{S}^j).\label{eq:Dip.Ham.}
\end{equation}

\par As explained in Sec.~\ref{secSU2}, the tetrahedral point group $\mathrm{T}$ is a decoupling group for this term. From table~\ref{tab2:point_group_factorizations}, we observe that it can be factorized as a product of two subgroups, $\mathrm{T}=\mathrm{C}_3\mathrm{D}_2$, where $\mathrm{D}_2$ is a four-element group that describes the symmetry of a rectangle -\ie, rotational symmetries of angle $\pi$ around three orthogonal axes-, and $\mathrm{C}_3$ is a three-element group that describes a rotational symmetry of angle $2\pi/3$ around an axis. 
\begin{figure}[t!]
    \centering
    \includegraphics[width=0.9\linewidth]{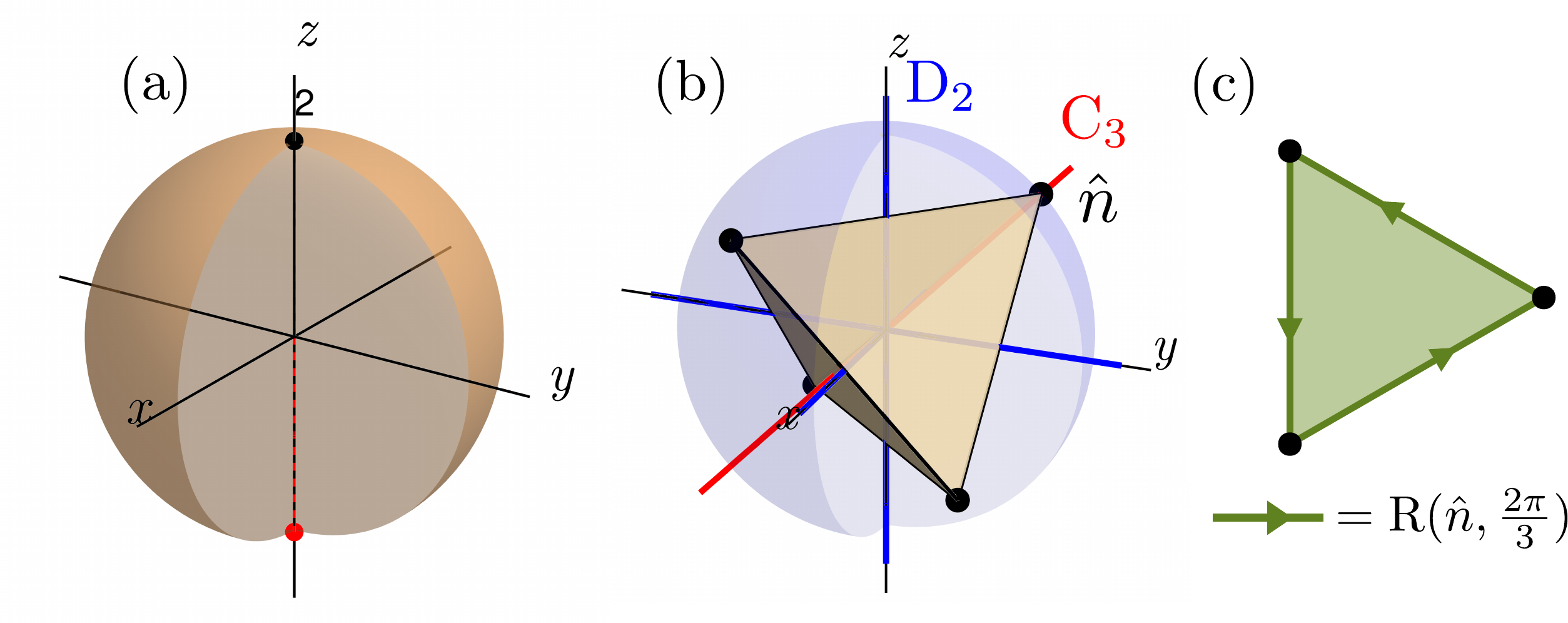}
    \caption{(a) Majorana constellation of the dipole-dipole Hamiltonian $H_{\mathrm{dd}}^{\mathrm{RWA}}$ in Eq.~\eqref{eq:Dip.Ham.}. (b) Tetrahedron whose $\mathrm{D}_2$ symmetry axes coincide with those of the Majorana constellation. (c) Cayley graph of the $\mathrm{C}_3$ symmetry group.}
    \label{fig4:Majorana.Hdd}
\end{figure}
\par In order to identify the rotational symmetries in the Hamiltonian $H_{\mathrm{dd}}^{\mathrm{RWA}}$, we can determine its Majorana constellation~\cite{Read2025platonicdynamical,Serrano_2020} and find that the operator has a $\mathrm{D}_{\infty}$ symmetry (and thus $\mathrm{D}_2$) about the $z$ axis [see Fig.~\ref{fig4:Majorana.Hdd}a]. Consequently, a $\mathrm{C}_3$ symmetrization along $\hat{n}$ is sufficient to suppress $H_{\mathrm{dd}}^{\mathrm{RWA}}$, which can be implemented \emph{via} the following three-pulse decoupling sequence  
\begin{equation}
\label{Eq.Seq.3P}
  \overset{\tau_0}{\mydash}~\qty(\frac{2\pi}{3})_{\hat{n}}\overset{\tau_0}{\mydash}~\qty(\frac{2\pi}{3})_{\hat{n}}\overset{\tau_0}{\mydash}~\qty(\frac{2\pi}{3})_{\hat{n}}\equiv  \left( \overset{\tau_0}{\mydash}~\qty(\frac{2\pi}{3})_{\hat{n}} \right)^{\times 3},
\end{equation}
where $\qty(\frac{2\pi}{3})_{\hat{n}}$ denotes a rotation of angle $2\pi/3$ around the axis $\hat{n} = \frac{1}{\sqrt{3}}\qty(-1,1,1)$ of duration $\tau_p$, $\times 3$ indicates that the pulse sequence in brackets must be applied three consecutive times, and $\tau=\tau_0+\tau_p$ is the pulse interval that includes the waiting time and the duration of the pulse [see Fig.~\ref{fig3:nesting}]. We emphasize that in this work, we use $\mathrm{R}(\hat{n},\theta)\in\mathrm{SO}(3)$ to denote the rotation of a spin system by an angle $\theta$ around the axis $\hat{n}$, and $(\theta)_{\hat{n}}\equiv e^{-i\theta \hat{n}\bcdot\mathbf{J}}$ to denote the corresponding $\mathrm{SU}(2)$ operator acting on the system Hilbert space.
\par The Hamiltonian actually possesses a $\mathrm{D}_{\infty}$ symmetry, so that the $\mathrm{C}_3$ symmetrization can be implemented around any axis $\mathrm{R}(\hat{z},\varphi)\hat{n}$ for an arbitrary angle $\varphi$. This rotated axis alters the orientation of the tetrahedral point group [see Fig.~\ref{fig4:Majorana.Hdd}b]. These are the only rotation axes for which the inherent symmetry of $H_{\mathrm{dd}}^{\mathrm{RWA}}$ can be exploited to effectively perform the $\mathrm{T}$-symmetrization operation.

\par Since this sequence naturally forms an Eulerian path in the Cayley graph of the $\mathrm{C}_3$ group [see Fig.~\ref{fig4:Majorana.Hdd}c], the symmetrization in the finite-pulse regime implements the quantum operation 
\begin{equation}
S \mapsto \Pi_{\mathrm{C}_3}\qty[F_{\Gamma_{\mathrm{C}_3}}^{(\tau)}\qty(S)]
\end{equation}
where $\Pi_{\mathrm{C}_3}$ is the relevant $\mathrm{C}_3$-symmetrization operation and $\Gamma_{\mathrm{C}_3} = \big\{e^{-i\frac{2\pi}{3}\hat{n}\bcdot\vec{J}}\big\}$ contains only one generator, with $\vec{J} = \sum_i\vec{S}^i$ the collective spin operator. In the case where $U = \big\{u(t) = e^{-i\int_0^t dt'\,f(t') \hat{n}\bcdot \vec{J}}\big\}$, the generator of $u(t)$ is $\hat{n} \bcdot \mathbf{J}$ and, consequently, $F^{(\tau)}_{\Gamma_{\mathrm{C}_3}}$ commutes with the symmetrization $\Pi_{\mathrm{C}_3}$. Thus, the decoupling sequence is robust to finite-duration pulses. Additional robustness to control errors might be achieved by appropriately shaping the control pulse profile~\cite{Guéron_1991,McDonald_1991,Goswami_2003,Warren_1988}, but the use of more complex composite pulses~\cite{Levitt_1986} is generally prohibited, as the robustness to finite duration errors is only satisfied when the generating pulse commutes with the group elements at all times. 

\par In the limiting case where each pulse occupies the full cycle time ($\tau_0=0$, i.e., no free-evolution interval), this scheme reduces to the well-known Lee-Goldburg (LG) sequence~\cite{LeeGoldburg_1965}, widely used for homonuclear dipolar decoupling in solid-state NMR spectroscopy~\cite{Mote_2016,Levitt_1993, Haeberlen_1968, Mehring_1972}. In this regime, the sequence effectively rotates the spin by an angle $2\pi$ around an axis tilted relative to the $z$ axis by the so-called "magic angle" $\acos(1/\sqrt{3})\approx 54.7^\circ$, which geometrically corresponds to the angle between the $C_3$ and $D_2$ symmetry axes of a regular tetrahedron. This classic decoupling protocol thus naturally appears as a special case of our general framework. Interestingly, the continuous irradiation of the sample by the LG sequence can be understood as a continuous DD sequence enforcing a $\mathrm{C}_{\infty}$ symmetry, which decouples the Hamiltonian~\eqref{eq:Dip.Ham.} because $\mathrm{C}_3$ is a subgroup of $\mathrm{C}_{\infty}$. An advantage of the 3-pulse variant~\eqref{Eq.Seq.3P} of the LG sequence in solid-state NMR is that the waiting time between the pulses can be used to observe the magnetization of the sample, as shown in Refs.~\cite{Haeberlen_1968,Mehring_1972}. From the discussion above, it is clear that the $3n$-pulse sequence ($n=1,2,\dots $) 
\begin{equation}
    \left( \overset{\tau_0}{\mydash}~\qty(\frac{2\pi}{3n})_{\hat{n}} \right)^{\times 3n},
\end{equation}
which implements a $\mathrm{C}_{3n}$ symmetrization, is an adequate decoupling sequence which offers more free-evolution intervals during which the magnetization can be measured.

\subsection{Dipole-dipole interaction and disorder}
We now consider a quantum system with dipole-dipole interactions~\eqref{eq:Dip.Ham.} and disorder under the RWA, described by
\begin{equation}
    H^{\mathrm{RWA}}_{\mathrm{dis+dd}} = \sum_i \delta_iS_z^i + \sum_{i<j}\Delta_{ij} \qty(3S_z^iS_z^j - \mathbf{S}^i \bcdot\mathbf{S}^j)\label{eq:Dip.Dis.Ham.}.
\end{equation}
While the shortest decoupling sequence for this Hamiltonian is given by the $6$-pulse  echo+WAHUHA sequence~\cite{Lukin_2020}, state-of-the-art sequences include the $24$-pulse yxx24~\cite{Cappellaro_2022} and 60-pulse DROID-$60$~\cite{Lukin_2020}, among others~\cite{Cory_1990, REV8, Mansfield_1973, Burum_1979}. These sequences are robust to both finite-duration and flip-angle (\ie, systematic over- and under-rotations) errors, and typically provide high-order decoupling in the Magnus expansion. 
\par
\begin{figure}[t!]
    \centering
    \includegraphics[width=0.9\linewidth]{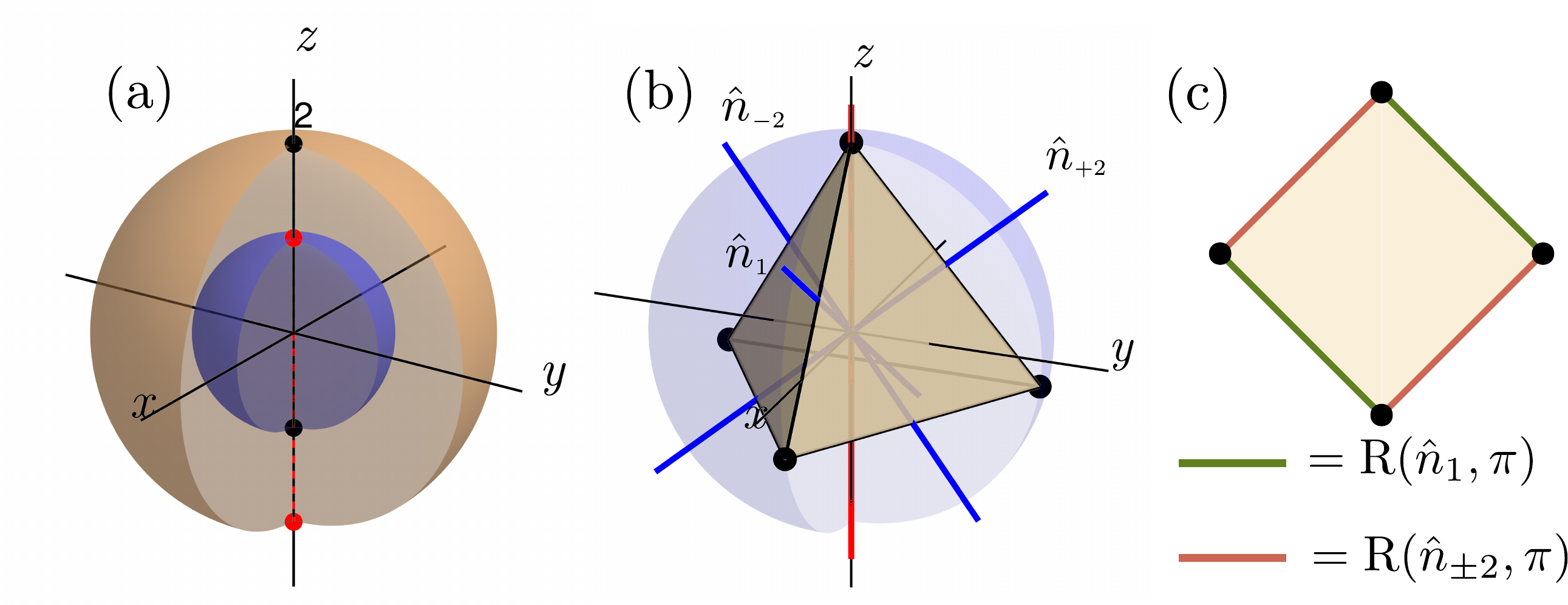}
    \caption{(a) Majorana constellations of the Hamiltonian \eqref{eq:Dip.Dis.Ham.}. (b) Tetrahedron whose $\mathrm{C}_3$ symmetry axis ($z$ axis) coincides with that of the Majorana constellation. The axes shown in blue are associated with the $\mathrm{D}_2$ symmetry. (c) Cayley graph of the $\mathrm{D}_2$ group. An undirected edge on a Cayley graph should be understood as two directed edges with opposite directions.}
    \label{fig5:MajoranaHddHdis}
\end{figure}

 The addition of the disorder term in \eqref{eq:Dip.Dis.Ham.} breaks the $\mathrm{D}_{\infty}$ symmetry of $H_{\mathrm{dd}}$. However, the interaction Hamiltonian still has a $\mathrm{C}_{\infty}$ symmetry about the $z$ axis that we can leverage. Through the factorization $\mathrm{T}=\mathrm{C}_3 \mathrm{D_2}$, we only have to apply the remaining $\mathrm{D}_2$ symmetrization about the corresponding axes shown in Fig.~\ref{fig5:MajoranaHddHdis}. This leads to a four-pulse sequence that cancels~\eqref{eq:Dip.Dis.Ham.} and is constructed from an Hamiltonian path on the Cayley graph of the $\mathrm{D}_2$ group, 
\begin{equation}
\label{Eq.4pul}
\left(     \overset{\tau_0}{\mydash}~(\pi)_{\hat{n}_1}~\overset{\tau_0}{\mydash}~(\pi)_{\hat{n}_{\pm 2}}
\right)^{\times 2}
,
\end{equation}
where $\hat{n}_1 = \frac{1}{\sqrt{3}}\qty(\sqrt{2},0,1)$ and the second axis is chosen from one of the two options $\hat{n}_{\pm 2}=\big(-\frac{1}{\sqrt{6}}, \pm\frac{1}{\sqrt{2}}, \frac{1}{\sqrt{3}}\big)$. Once the second axis has been selected, it must be used for the entire Eulerian sequence. In fact, note that the $\mathrm{C}_{\infty}$ symmetry means that any other pair of axes $\qty{\mathrm{R}(\hat{z},\varphi)\hat{n}_1, \mathrm{R}(\hat{z},\varphi)\hat{n}_{\pm2}}$ ($\forall \varphi$) can be chosen. Note that for this sequence the $\pi$ rotations can be implemented clockwise or counter clockwise as this does not change the Cayley graph, which leaves more degrees of freedom. In the ideal pulse regime, \ie, when the pulses can be considered infinitely short and strong, the sense of rotation does not change the sequence properties.

\par The pulse sequence~\eqref{Eq.4pul} is the shortest known DD sequence which decouples the Hamiltonian~\eqref{eq:Dip.Dis.Ham.}, with only four pulses. Moreover, since $\mathrm{D}_2$ is an inaccessible symmetry for the entire $\BHs^{(1)}$ irrep (see Table~\ref{tab1:Point.Groups}), the sequence actually corrects for arbitrary disorder $H_{\mathrm{dis}} = \sum_i \delta_i\, \hat{m}_i \bcdot\vec{S}^i$.

\par For additional robustness properties, one can design an eight-pulse Eulerian sequence on the same Cayley graph, given by 
\begin{equation}
     \qty(\overset{\tau_0}{\mydash}~(\pi)_{\hat{n}_1}~\overset{\tau_0}{\mydash}~(\pi)_{-\hat{n}_{\pm 2}})^{\cross 2} \qty(\overset{\tau_0}{\mydash}~(\pi)_{-\hat{n}_{\pm 2}}~\overset{\tau_0}{\mydash}~(\pi)_{\hat{n}_{1}})^{\cross 2} .
     \label{TEDDY}
\end{equation}
We refer to this sequence as $\mathrm{TEDDY}$, as it is a promising shorter alternative to the $\mathrm{TEDD}$ sequence (with 24 pulses) in systems where the RWA is valid, using three times fewer pulses. For this sequence, we choose the pulses such that it implements a counterclockwise rotation around one of the axes $\hat{n}_{\pm 2}$ (or, equivalently, a clockwise rotation around $-\hat{n}_{\pm 2}$), as we find that the corresponding sequence offers better decoupling properties on the disorder term. Due to its Eulerian design, the sequence implements the following quantum operation 
\begin{equation}
    S \mapsto \frac{1}{2}\sum_{\lambda=1,2}\Pi_{\mathrm{D}_2}\qty(F^{(\Gamma_{\mathrm{D}_2},\tau)}_{\lambda}(S))
\end{equation}
where 
\begin{equation}
    F^{(\Gamma_{\mathrm{D}_2},\tau)}_{\lambda}(S) = \frac{1}{\tau} \int_0^{\tau} u_{\lambda}^{\dagger}(t)\, S\, u_{\lambda}(t)dt
\end{equation}
with $U = \qty{u_1(t),u_2(t)}$ such that $u_1(\tau) = e^{-i\pi\hat{n}_1\bcdot \vec{J}}$ and $u_2(\tau) = e^{i\pi\hat{n}_{\pm 2}\bcdot \vec{J}}$. In the case where
\begin{equation}
    \begin{cases}
        u_1(t) &= e^{-i\int_0^tf_1(t')dt' \hat{n}_1\bcdot\vec{J}} \\[4pt] 
        u_{2}(t) &= e^{i\int_0^tf_2(t')dt' \hat{n}_{\pm 2}\bcdot\vec{J}}
    \end{cases},
\end{equation}
and if the control profiles are time-symmetric, \ie, $f_{\lambda}(t) = f_{\lambda}(\tau-t)$ for $\lambda=1,2$ and $\forall~t\in[0,\tau]$, it is possible to prove that robustness to errors caused by finite-duration pulses is guaranteed (see Appendix~\ref{Ap.Robustness} for the proof). Due to the restriction imposed on the shape of the control profile, there is little freedom left to exploit to increase robustness to control errors through pulse shaping or composite pulse techniques. However, thanks to the Eulerian design, the sequence~\eqref{TEDDY} is robust to any systematic control error for which $\mathrm{D}_2$ is a decoupling group, in that these errors are suppressed to first order in the Magnus expansion. This includes many relevant pulse errors such as systematic over- and under-rotation, axis misspecification or errors in the shape of the control profile. More details on the robustness analysis can be found in the Appendix~\ref{Ap.Robustness}. 

\par The decoupling performance of our sequence can be readily enhanced by imposing reflection symmetry~\cite{Cappellaro_2022, Lukin_2020, Levitt_2007,Preskill_2011}. This is achieved by appending the original pulse sequence~\eqref{TEDDY} with its time-reversed, sign-inverted counterpart—constructed by reversing the order of the pulses in~\eqref{TEDDY} and flipping the sign of the control profiles. This simple construction yields a 16-pulse sequence that cancels the Hamiltonian~\eqref{eq:Dip.Dis.Ham.} up to second order in the Magnus expansion, even within the finite-pulse regime. 
\begin{figure}[hbt]
    \centering
    \includegraphics[width=0.95\linewidth]{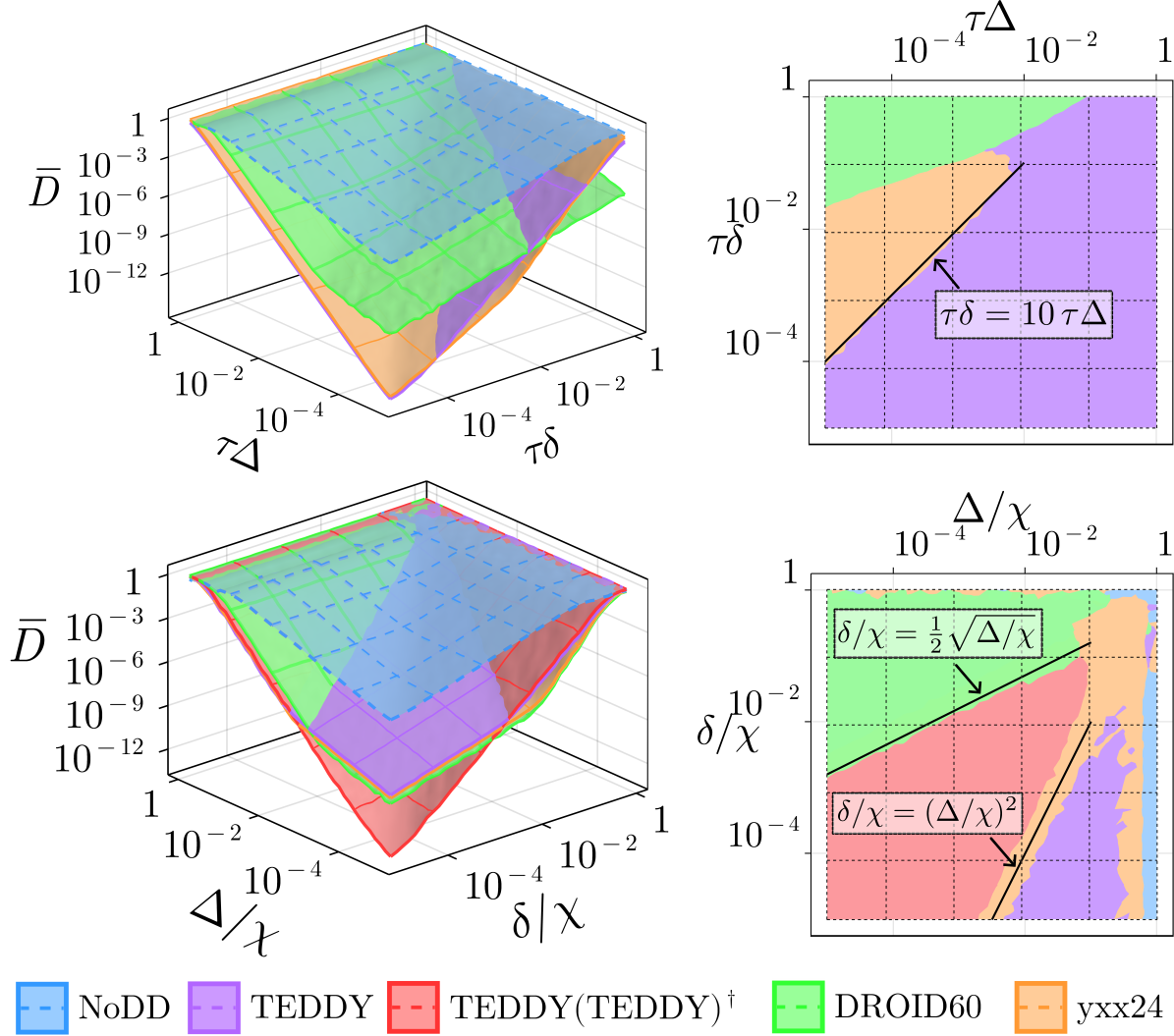}
    \caption{Left: average distance between the identity and the noisy propagator of an ensemble of interacting qubits described by the Hamiltonian~\eqref{eq:Dip.Dis.Ham.} under the application of different DD sequences, in the ideal (top, $\tau_p=0$) and finite (bottom, $\tau_0=0$) pulse regimes. $\mathrm{NoDD}$ corresponds to a free evolution of a duration equal to that of the shortest sequence. Right: map showing the best protocol for each point in parameter space. The sequence $\mathrm{(TEDDY)(TEDDY)^{\dagger}}$ is not shown in the ideal pulse regime, as its performance is not superior to that of $\mathrm{TEDDY}$.}
    \label{fig:6}
\end{figure}
In order to compare performance with that of state-of-the-art sequences, we compute the distance
\begin{equation}
    D(\mathds{1}_S,U)=\sqrt{1-\frac{1}{2^N}\abs{\tr[U]}}
\end{equation}
between the identity operator and the noisy evolution operator $U$ of a system of $N$ interacting qubits with the Hamiltonian~\eqref{eq:Dip.Dis.Ham.} on which we apply a DD sequence. We consider two limiting regimes: the ideal pulse regime (pulses are instantaneous and separated by a pulse interval $\tau$) and the finite pulse regime (where there is no waiting time and the duration of the pulses are set by the pulse amplitude $\chi$, so that the pulse interval is $\tau = \theta/\chi$ where $\theta$ is the rotation angle of the DD pulse). Figure~\ref{fig:6} shows the average distance $\bar{D}$ in the parameter space $(\tau\delta, \tau\Delta)$ and $(\delta/\chi, \Delta/\chi)$, where $\delta = \norm{H_{\mathrm{dis}}}$ and $\Delta=\norm{H_{\mathrm{dd}}}$. The average is made from a sample of 20 Hamiltonians characterized by a set of randomly generated frequencies $\qty{\delta_i,\Delta_i}$, according to a uniform distribution. We compare our sequences with DROID-60~\cite{Lukin_2020} and yxx24~\cite{Cappellaro_2022}, which perform best in the disorder-dominated and interaction-dominated regime respectively. In the ideal pulse regime, we can see that $\mathrm{TEDDY}$ slightly outperforms both sequences in the interaction-dominated region of the parameter space, where the data indicates that our 8-pulse sequence achieves second order of decoupling. In the finite-pulse regime, numerical evidence shows that $\mathrm{TEDDY}$ outperforms all sequences in the regime of parameters $\left(\Delta/\chi\right)^2 \gtrsim \delta/\chi \gtrsim \frac{1}{2} \sqrt{\Delta / \chi} $, i.e., when the dipole-dipole interaction strength far exceeds the amplitude of the disorder. In a vast region of the parameter space, the 16-pulse sequence (referred to as $\mathrm{(TEDDY)(TEDDY)^{\dagger}}$) drastically outperforms all sequences as it is the only one which cancels all second-order terms when pulses are not instantaneous.

\subsection{\texorpdfstring{$K$}{Lg}-body multilinear interactions}

\begin{figure}[t]
    \centering
    \includegraphics[width=0.8\linewidth]{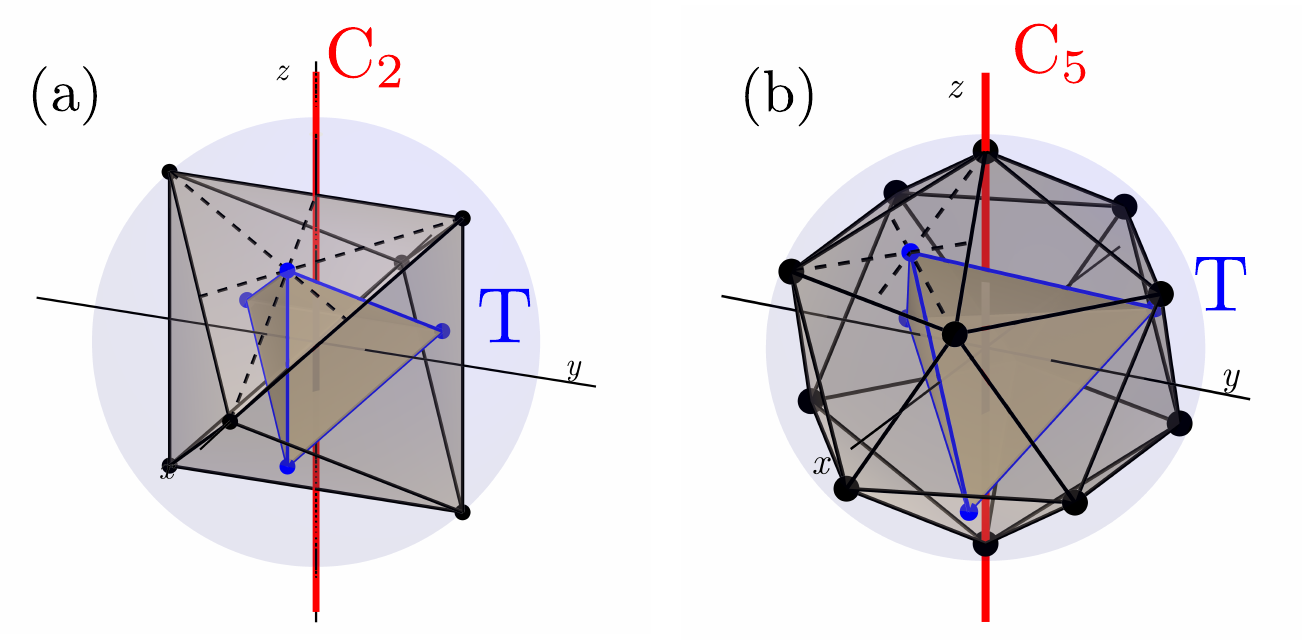}
    \caption{Representation of $\mathrm{T}$ symmetry in (a) an octahedron and (b) an icosahedron, with $\mathrm{C}_2$ symmetry about the $z$ axis in (a) and $\mathrm{C}_5$ symmetry about the same axis in (b).}
    \label{fig7:O&I}
\end{figure}

More exotic anisotropic $K$-body multilinear interactions\footnote{By $K$-body multilinear, we mean that the interaction Hamiltonian can be expressed as a sum of $K$-fold tensor products of operators linear in the spin operators $\qty(S_x,S_y,S_z)$. A multilinear interaction is anisotropic if it does not have an isotropic component, which is invariant under global $\mathrm{SU}(2)$. For example, an isotropic trilinear interaction between spins $(s_1,s_2,s_3)$ is $\vec{S}^1\bcdot\qty(\vec{S}^2\cross \vec{S}^3)$.}~\cite{Büchler_2007,Mezzacapo_2014} can also be suppressed by the Platonic decoupling groups, as long as $K\leq 5$. For example, the 24-element octahedral point group is a decoupling group for three-body multilinear interactions, while the 60-element icosahedral point group decouples both four- and five-body multilinear interactions~\cite{Read2025platonicdynamical}. When the RWA is valid, these interactions are invariant under rotation around the $z$ axis and this $\mathrm{C}_{\infty}$ symmetry can be leveraged by the point group factorizations $\mathrm{O}=\mathrm{C}_2\mathrm{T}$ and $\mathrm{I}=\mathrm{C}_5\mathrm{T}$ (see Table~\ref{tab2:point_group_factorizations}). Therefore, the $\mathrm{T}$ symmetrization in the orientation shown in Fig.~\ref{fig7:O&I} is sufficient to enforce an octahedral or icosahedral symmetry, thus suppressing anisotropic $K$-body multilinear interactions with $K\leq 3$ or $K\leq 5$, respectively.

\par
\begin{figure}[t]
    \centering
    \includegraphics[width=0.95\linewidth]{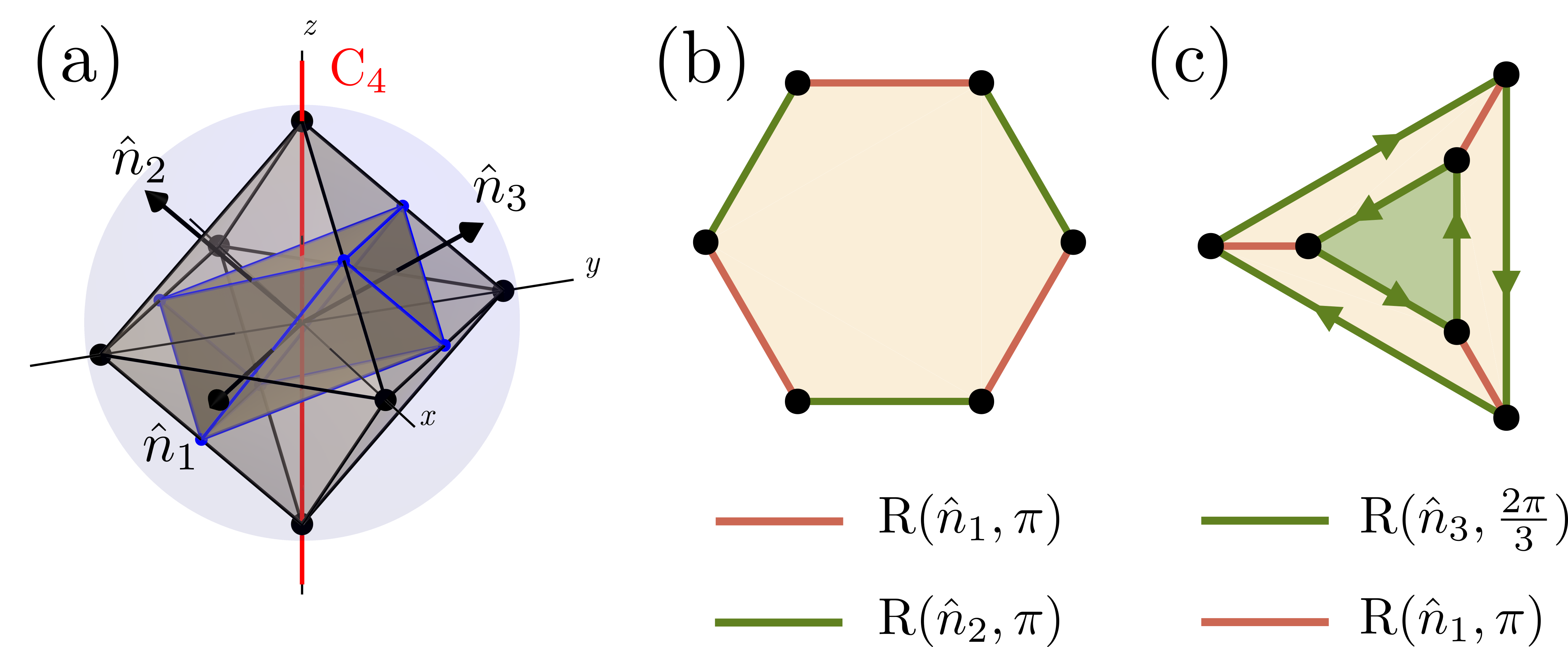}
    \caption{(a) Representation of the $\mathrm{D}_3$ symmetry in an octahedron with $\mathrm{C}_4$ symmetry about the $z$ axis. (b) and (c) Cayley graph of the $\mathrm{D}_3$ symmetry group with respect to the generating sets $\qty{2C_2}$ and $\qty{C_2 , C_3}$ respectively.}
    \label{fig8:D3}
\end{figure}

 The octahedral point group has an additional factorization $\mathrm{O}=\mathrm{C}_4\mathrm{D}_3$, where $\mathrm{D}_3$ describes the symmetry group of a triangular antiprism. We can thus take advantage of the existing $\mathrm{C}_{\infty}$ symmetry (and thus $\mathrm{C}_4$), and construct a sequence on the $\mathrm{D}_3$ Cayley graph. There are two distinct Cayley graph for this point group, depending on the generators chosen. For a generator set consisting of two $C_2$ elements (two rotations of angle $\pi$ about nonorthogonal axes), the Cayley graph has a ring shape, as shown in Fig.~\ref{fig8:D3}b, and the corresponding six-pulse DD sequence is given by 
\begin{equation}
        \qty(\overset{\tau_0}{\mydash}~(\pi)_{\hat{n}_1}~ \overset{\tau_0}{\mydash}~(\pi)_{\hat{n}_2} )^{\cross 3}\label{eq:D3a}
\end{equation}
where $\hat{n}_1=\frac{1}{\sqrt{2}}\qty(1,-1,0)$ and $\hat{n}_2=\frac{1}{\sqrt{2}}\qty(0,-1,1)$. Choosing a $C_2$ and $C_3$ element instead (rotations of angle $\pi$ and $2\pi/3$ respectively about orthogonal axes), we find the Cayley graph depicted in Fig.~\ref{fig8:D3}c and the pulse sequence
\begin{equation}
         \qty(\overset{\tau_0}{\mydash}~\qty(\frac{2\pi}{3})_{\hat{n}_3}~\overset{\tau_0}{\mydash}~\qty(\frac{2\pi}{3})_{\hat{n}_3}~\overset{\tau_0}{\mydash}~(\pi)_{\hat{n}_1})^{\cross 2}\label{eq:D3b}
\end{equation}
where $\hat{n}_3=\frac{1}{\sqrt{3}}\qty(1,1,1)$. 

\subsection{Decoupling of a dephasing qudit}

Let us consider a single qudit, with $d$-level Hamiltonian
\begin{equation}
    H_S = \sum_{n = 0}^{d-1} E_n \ket{n}\bra{n},
\end{equation}
interacting with a bosonic bath with Hamiltonian $H_B$ and system-bath coupling Hamiltonian $H_{SB}$ given by 
\begin{equation}
    \begin{aligned}
        H_B &= \sum_k \omega_k b_k^{\dagger}b_k,\\ 
        H_{SB} &= S\otimes \sum_k \qty(g_kb_k^{\dagger} + g_k^*b_k) ,
    \end{aligned}
\end{equation}
for some system operator $S$, where $g_k$ describes the coupling strength between the qudit and the $k$th mode of the bosonic field of frequency $\omega_k$. In the rotating frame with respect to $H_S+H_B$, the time-dependent coupling Hamiltonian is given by 
\begin{equation}
    H_{SB}(t) = e^{iH_St}S e^{-iH_St}\otimes\sum_k\qty(g_kb_k^{\dagger}e^{-i\omega_kt} + g_k^*b_ke^{i\omega_kt}).
\end{equation}
In the case of a dephasing dynamics for the qudit, the system operator $S$ commutes with $H_S$ and is diagonal in the eigenstates basis $\{\ket{n}\}$, so that the coupling Hamiltonian reduces to 
\begin{equation}
        H_{SB}(t) = S\otimes\sum_k\qty(g_kb_k^{\dagger}e^{-i\omega_kt} + g_k^*b_ke^{i\omega_kt}).\label{eq:inter.boson.}
\end{equation}
To suppress this coupling, we can send a DD sequence composed of $\mathrm{SU}(2)$ pulses in the rotating frame, provided that the control Hamiltonian 
\begin{equation}
    H_{\mathrm{DD}}(t)  = \chi(t) \, \hat{n}(t)\bcdot \vec{J}
\end{equation}
can be implemented in the rotating frame. Here $\vec{J} = \qty(J_x,J_y, J_z)$ denotes the angular momentum operators of a fictitious spin-$j$, with $j = \frac{d-1}{2}$. The most general diagonal operator $S$ for a $d$-level system belongs to the irreps decomposition 
\begin{equation}
    S \in \bigoplus_{L = 0}^{d-1}\BHs^{(L)}\label{Lprime decomposition}
\end{equation}
so that we look for $\mathrm{SU}(2)$ symmetries that are inaccessible to all $L$-irreps with $L \leq d-1$. As shown in Table~\ref{tab1:Point.Groups}, this condition can only be satisfied for qudits with $d < 7$, since there are no inaccessible point groups to the irrep $\BHs^{(6)}$. Even for qudits with these numbers of levels, the order of the inaccessible point groups can be very large, which limits the practical use of the decoupling scheme. However, because $S$ is diagonal, it is invariant under rotations about the $z$ axis and this symmetry can be leveraged to more effectively suppress dephasing~; this leaves us with smaller decoupling groups which are listed in Table~\ref{tab3:dephasing}. Although the numbers of pulses have been significantly reduced, it is worth pointing out that shorter decoupling sequences have been constructed, which suppress dephasing in a single qudit~\cite{Tripathi_2024,iiyama_2024}. Note that if $S$ has support on a subset of the irreps in the decomposition~\eqref{Lprime decomposition}, it may be possible to further reduce the decoupling group. For example, if the dephasing operator $S$ of a qudit with $d=6$ levels of energy has no support on the $L=4$-irrep, then an octahedral symmetrization is sufficient to achieve decoupling and the $\mathrm{D}_3$ sequence can be used instead of a tetrahedral sequence, which further reduces the number of pulses by a factor of two.

\begin{table}[t]
    \setlength{\tabcolsep}{8pt}
    \renewcommand*{\arraystretch}{1.5}
    \centering
    \begin{tabular}{c|c|c|c|c}
       \multirow{2}{*}{$d$}  & \multicolumn{2}{c|}{Universal decoupling} & \multicolumn{2}{c}{Dephasing only}  \\\cline{2-5}
         & Point group & Order & Point group & Order \\ \hline \hline 
         2 & $\mathrm{D}_2$ & 4 & $\mathrm{C}_2$ & 2 \\ 
         3 & $\mathrm{T}$ & 12 & $\mathrm{D}_2$ & 4 \\ 
         4 & $\mathrm{O}$ & 24 & $\mathrm{D}_3$ & 6 \\ 
         5,\,6 & $\mathrm{I}$ & 60 & $\mathrm{T}$ & 12 \\ 
    \end{tabular}
    \caption{Decoupling groups for both universal decoupling and dephasing-only decoupling for all qudits up to $d=6$ levels. For the dephasing qubit ($d=2$), the $\mathrm{C}_2$ decoupling group leads to the CPMG (spin-echo) sequence~\cite{Viola_1999}. The orientations of the relevant $\mathrm{D}_2$, $\mathrm{D}_3$ and $\mathrm{T}$ symmetry groups for dephasing-only decoupling correspond to those shown in Figs.~\ref{fig5:MajoranaHddHdis}b, \ref{fig8:D3}a and~\ref{fig7:O&I}b respectively.}
    \label{tab3:dephasing}
\end{table}

\par Although Eulerian sequences can be constructed for each of these decoupling groups, it should be emphasized that robustness to all finite durations is not guaranteed because the finite-duration errors do not necessarily have the same symmetry as the dephasing Hamiltonian, as mentioned in Sec.~\ref{Sec.fin.dur.lev}. However, using pulse shaping or composite pulse techniques, it may be possible to engineer the finite-duration error Hamiltonian such that the necessary rotational invariance about the $z$ axis appears, which should guarantee robustness to finite-duration errors at the cost of more complex generating pulses. The robustness to control errors provided by the Eulerian design would then ensure good sequence performances in the case where these pulses are more prone to errors.

\par
\begin{figure}[t]
    \centering
    \includegraphics[width=\linewidth]{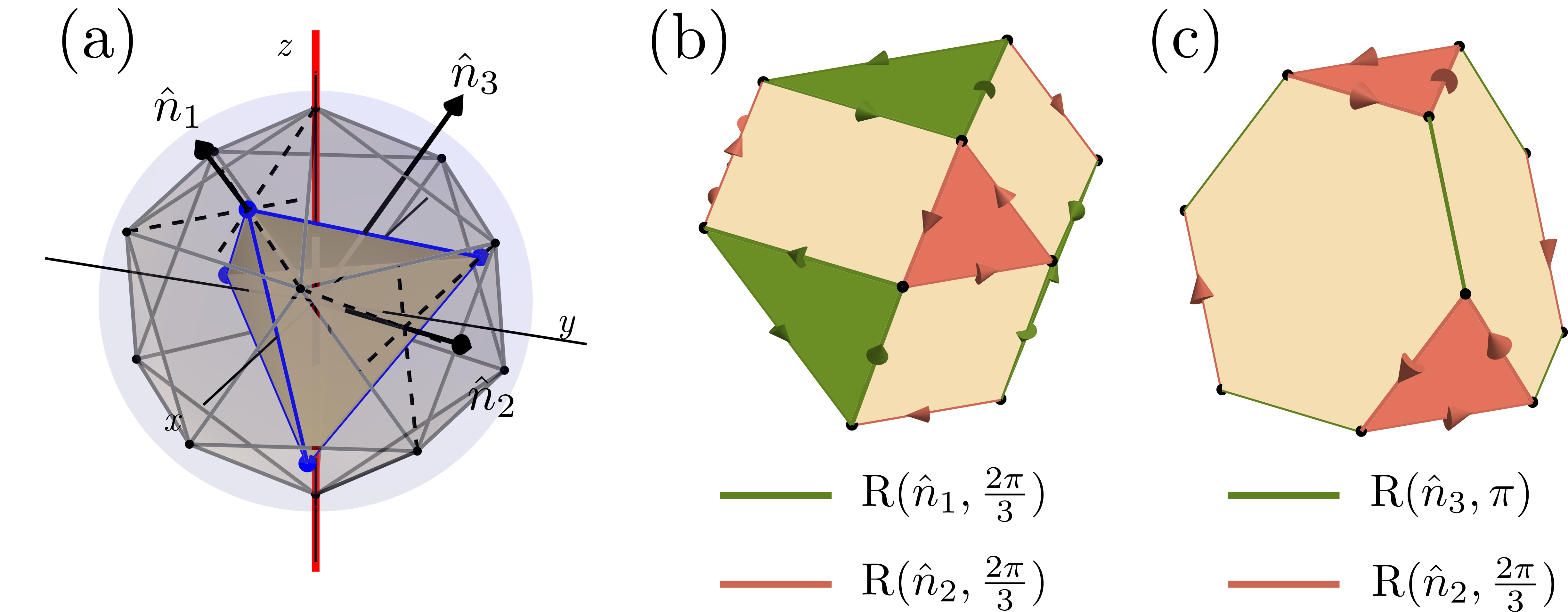}
    \caption{(a) Representation of a $\mathrm{T}$ symmetry inside an icosahedron with a $\mathrm{C}_5$ symmetry on the $z$ axis. (b) (resp.~(c)) Cayley graph of the tetrahedral point group with respect to a generating set $\qty{2C_3}$ (resp.~$\qty{C_3,C_2}$). Note that the two $C_3$ axes have been chosen to satisfy the defining relation $\qty[\mathrm{R}(\hat{n}_1,\frac{2\pi}{3})\mathrm{R}(\hat{n}_2,\frac{2\pi}{3})]^2=E$ where $E$ is the identity.}
    \label{fig9:TinI}
\end{figure}

 As a brief illustration of the formalism, we consider the dephasing of a spin-$2$ tetrahedron state, a fragile anticoherent state to order 2 useful in rotation metrology~\cite{2018Goldberg,Denis_2022,2024Ferretti}. For a certain orientation of the axes, this state is given by
\begin{equation}
    \ket{\psi} = c_1\ket{2,2} + c_2\ket{2,0} + c_1\ket{2,-2}
\end{equation}
with 
\begin{equation}
    \begin{aligned}
        c_1 &= \frac{-1 + \sqrt{2}i}{2\sqrt{3}}, &
        c_2 &= \frac{\sqrt{2} + i}{\sqrt{6}}.
    \end{aligned}
\end{equation}
The operator $S$ in the coupling Hamiltonian can, in all generality, be written as 
\begin{equation}\label{jumpTLM}
    S=\sum_{L=1}^{4}\sum_{M=-L}^{L} \omega_{L M}T_{L M}
\end{equation}
where $\qty{T_{L M}\,:\,M=-L,-L+1,\dots , L}$ are the multipolar operators~\cite{Varshalovich_1988}. 
The linear combination of $T_{L M}$'s  with a fixed $L$ is associated to a bi-colored Majorana constellation with $2L$ points~\cite{Serrano_2020,Read2025platonicdynamical}.
It was shown in Ref.~\cite{Read2025platonicdynamical} that the constellations with $\mathrm{C}_{\infty}$ symmetry about the $z$ axis are those of operators proportional to $T_{L 0}$. Therefore, an operator $S$ that is diagonal (invariant under rotation about the quantization axis $z$) can be written as $S=\sum_{L=1}^{4}\omega_{L0}T_{L0}$. According to Table~\ref{tab3:dephasing}, an operational DD sequence to decouple dephasing in this case corresponds to an oriented tetrahedral symmetry as in Fig~\ref{fig9:TinI}a. There are again two distinct Cayley graphs for the tetrahedral point group depending on the choice of generators. For a set of generators consisting of two $C_3$ elements (resp.\ one $C_3$ element and one $C_2$ element), the Cayley graph is shown in Fig.~\ref{fig9:TinI}b (resp.\ Fig.~\ref{fig9:TinI}c). Note that Cayley graphs are represented in three dimensions, but there is an equivalent two-dimensional representation that can be used to design DD sequences more easily. A valid choice (among many others) for the rotation axes is
\begin{equation}\begin{aligned}
    \hat{n}_1 &= R\left(-\hat{z},\frac{4\pi}{5}\right)\,R\left(-\hat{x} , \arccos\frac{\varphi}{\sqrt{\varphi+2}}\right)\,\begin{pmatrix}
        1-\varphi \\ 0 \\ \varphi
    \end{pmatrix}/\sqrt{3} \\ 
    \hat{n}_2 &= R\left(\hat{n}_3,\frac{\pi}{2}\right)\,\hat{n}_1 \\
    \hat{n}_3 &= R\left(-\hat{x} , \arccos\frac{\varphi}{\sqrt{\varphi+2}}\right)\,\hat{z}
\end{aligned}
\label{Eq.Axes.4.5}
\end{equation}
where $R(\hat{e},\theta)$ is the $\mathrm{SO}(3)$ matrix representing a rotation of angle $\theta$ about an axis $\hat{e}$ and $\varphi = \frac{1+\sqrt{5}}{2}$ is the golden ratio. Several DD sequences can be constructed as there are many valid Hamiltonian paths on the different Cayley graphs. Two such paths are given by 
\begin{equation}
\begin{aligned}
\mathrm{TDD}_1 ={}&
\mathrel{\overset{\tau_0}{\mydash}}~a~
\mathrel{\overset{\tau_0}{\mydash}}~b~
\mathrel{\overset{\tau_0}{\mydash}}~a~
\mathrel{\overset{\tau_0}{\mydash}}~a~
\mathrel{\overset{\tau_0}{\mydash}}~\overline{b}~
\mathrel{\overset{\tau_0}{\mydash}}~\overline{a} \\
&\mathrel{\overset{\tau_0}{\mydash}}~\overline{a}~
\mathrel{\overset{\tau_0}{\mydash}}~\overline{b}~
\mathrel{\overset{\tau_0}{\mydash}}~\overline{b}~
\mathrel{\overset{\tau_0}{\mydash}}~a~
\mathrel{\overset{\tau_0}{\mydash}}~a~
\mathrel{\overset{\tau_0}{\mydash}}~\overline{b}
\end{aligned}\label{TDD1}
\end{equation}
and
\begin{equation}
        \mathrm{TDD_2} = \qty(\overset{\tau_0}{\mydash}~c~\overset{\tau_0}{\mydash}~b~\overset{\tau_0}{\mydash}~b~\overset{\tau_0}{\mydash}~c~\overset{\tau_0}{\mydash}~\overline{b}~\overset{\tau_0}{\mydash}~\overline{b} )^{\cross 2}\label{TDD2}
\end{equation}
where 
\begin{equation}
    \begin{aligned}
        a &{}= \qty(\frac{2\pi}{3})_{\hat{n}_1}  & b &{}= \qty(\frac{2\pi}{3})_{\hat{n}_2} & c &{}= (\pi)_{\hat{n}_3}\\
        \overline{a} &{}= \qty(\frac{2\pi}{3})_{-\hat{n}_1} & \overline{b} &{}= \qty(\frac{2\pi}{3})_{-\hat{n}_2}&&
    \end{aligned}.
\label{Eq.Axes.4.5.body}
\end{equation}

\par
\begin{figure}[h]
    \centering
    \includegraphics[width=\linewidth]{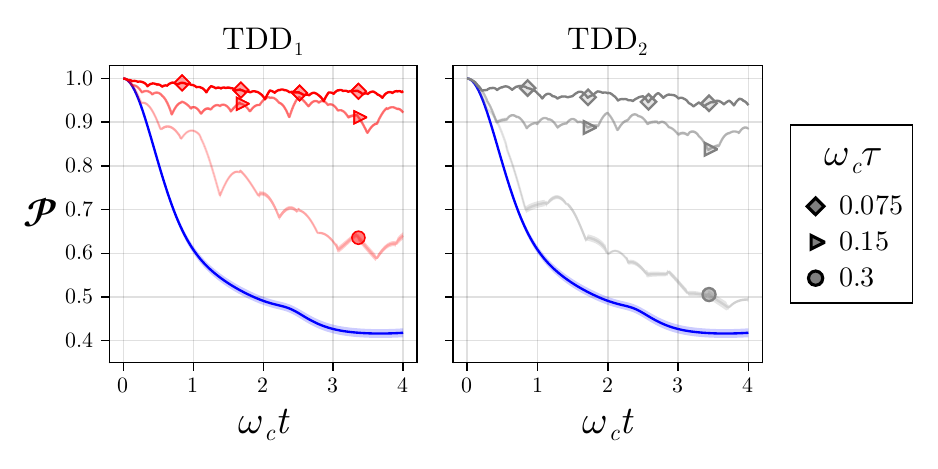}
    \caption{Purity as a function of time for the dephasing spin-2 tetrahedron state under no DD (blue lines), the $\mathrm{TDD_1}$ (red lines) and the $\mathrm{TDD_2}$ sequences (grey lines) for several values of the pulse interval $\omega_c\tau \in \qty{0.075,0.15,0.3}$. The purity at the refocusing time, \ie, after a complete iteration of the DD sequence, is indicated each time by a marker. Simulations were performed using the HOPS method and averaged over 500 quantum trajectories.}
    \label{fig10:Purity}
\end{figure}

 We model the coupling of the spin with the bosonic bath by a Lorentzian spectral density
\begin{equation}
    J(\omega) = \frac{g\kappa}{2}\frac{\kappa}{\kappa^2 + (\omega - \omega_c)^2}.
\end{equation}
In our simulations, we set $\kappa = \omega_c/10$ and $g=\omega_c$. We compute the purity of the evolved state, $\mathcal{P}=\Tr[\rho^2]$, as a function of $\omega_c t$ without DD and with the application of $\mathrm{TDD}_{1}$ or $\mathrm{TDD}_{2}$, using the Hierarchy of Pure States (HOPS) method~\cite{Suess_2014}. The dephasing operator \eqref{jumpTLM} is chosen with frequencies $\omega_{L 0}\equiv\omega$ $\forall\, L$, where $\omega$ is chosen such that $\norm{S} = g$, to ensure that the different multipolar operators contribute equally. It should be noted that this decoherence model is a simplified and illustrative model specifically designed to ensure that constellations with $1$ to $4$ pairs of antipodal stars are represented. We fix the waiting time between two pulses to $\tau_0 = 0.8\tau$, and the duration of the $\pi$ and $2\pi/3$ pulses to $\tau_{\pi}=0.2\tau$ and $\tau_{2\pi/3}=\frac{2}{3}0.2\tau$ for various values of the parameter $\omega_c\tau \in \qty{0.075,0.15,0.3}$, so that the duration of each pulse takes at most 20\% of the total pulse interval\footnote{Note that in this case, the pulse interval varies for $\pi$ and $2\pi/3$ pulses, but the waiting time between pulses does not.}. The results are plotted in Fig.~\ref{fig10:Purity} and show that a significant increase in coherence time can be achieved by repeatedly applying the pulse sequence, particularly when the time interval $\tau$ is short.

\subsection{Hierarchical decoupling in spin qudits} 
Relevant quantum platforms proposed for quantum computing make use of qudits encoded into atomic~\cite{omanakuttan_2021,omanakuttan_2024phd,Omanakuttan_2024} or solid-state~\cite{Onizhuk_2024,Xi_2025} spin qudits. In such (potentially large) spins, noise typically occurs due to interactions with a fluctuating microscopic magnetic field, leading to depolarization and dephasing~\cite{Denis_2022}. These errors are linear in the spin operators and belong to the 1-irrep $\BHs^{(1)}$. However, higher-order errors can also occur due, for example, to quadrupole interactions with an electric field~\cite{Omanakuttan_2024, Lukin_2024, Onizhuk_2024} or strain inhomogeneities~\cite{Lukin_2024,sarkar_2025,waeber_2019,chekhovich_2015}. These particular errors are quadratic in the spin operators and thus have support on the 2-irrep $\BHs^{(2)}$. In recent years, the possibility of encoding a qubit in a spin qudit was investigated, and quantum error correction codes were constructed to protect the logical qubit from such errors~\cite{Omanakuttan_2024, Gross_2021,Lim_2023, OmanakuttanAndGross_2023,Gross_2024}.

\par Both linear and quadratic errors can be suppressed by a tetrahedral symmetrization, as the tetrahedral point group is inaccessible to both $\BHs^{(1)}$ and $\BHs^{(2)}$. However, magnetic field noise can already be suppressed by a $\mathrm{D}_2$ symmetrization (see Table~\ref{tab1:Point.Groups}). In the case where linear errors in the spin operators (e.g.\ depolarization) are dominant with respect to quadratic errors (e.g.\ strain noise), a $\mathrm{D}_2$ symmetrization on a shorter timescale might be advantageous over the implementation of the full $\mathrm{T}$ sequence. This can be performed by nesting the relevant four-pulses sequence into a $\mathrm{C}_3$ outer-layer, exploiting the factorization $\mathrm{T}=\mathrm{C}_3 \mathrm{D}_2$. A sequence that performs $\mathrm{C}_3\qty[\mathrm{D}_2]$ is given by
\begin{equation}
\qty( \overset{\tau_0}{\mydash}~(\pi)_{\hat{n}_1}~\overset{\tau_0}{\mydash}~(\pi)_{\hat{n}_{ 2}}~\overset{\tau_0}{\mydash}~(\pi)_{\hat{n}_1}~\overset{\tau_0}{\mydash}~(\pi)_{\hat{n}_{2}}~\qty(\frac{2\pi}{3})_{\hat{n}_3})^{\cross 3}\label{eq.D2C3}
\end{equation}
with $\hat{n}_1=\hat{x}$, $\hat{n}_2=\hat{y}$ and $\hat{n}_3=\frac{1}{\sqrt{3}}\qty(1,1,1)$. We emphasize that the last two pulses in the bracket should be applied without any waiting time in between. Alternatively, one could also replace them by a single pulse implementing the same operation as these two successive rotations~; this pulse will also represent a rotation of the tetrahedral group for which the angle $\theta_{23}$ and axis $\hat{n}_{23}$ can be easily determined~\cite{Levitt_1986,Counsell_1985}. In the case of a $\pi$-pulse followed by a $2\pi/3$-pulse and for any $\hat{n}_2$ and $\hat{n}_3$, they are found by solving the set of equations
\begin{equation}
    \begin{aligned}
        \cos\qty(\frac{\theta_{23}}{2}) &= -\frac{\sqrt{3}}{2}\hat{n}_2\boldsymbol{\cdot}\hat{n}_3 \\
        \sin\qty(\frac{\theta_{23}}{2}) \hat{n}_{23}&= \frac{1}{2}\hat{n}_1 - \frac{\sqrt{3}}{2}\hat{n}_1\cross\hat{n}_2
    \end{aligned}
\end{equation}
For the $\hat{n}_2$ and $\hat{n}_3$ specified above, we find $\theta_{23} = \frac{2\pi}{3}$ and $\hat{n}_{23} = \frac{1}{\sqrt{3}}\qty(1,\,-1,\,-1)$.
\par Additional cubic errors in the spin operators can be corrected on a longer timescale by concatenating the above sequence with the relevant two-pulse sequence to implement the octahedral symmetrization,
\begin{equation}\mathrm{C}_2\qty[\mathrm{C}_3\qty[\mathrm{D}_2]]\equiv \qty(\mathrm{C}_3\qty[\mathrm{D}_2]\,(\pi)_{\hat{n}_4})^{\cross 2} ,
\label{Eq.O.hierarchy}
\end{equation}
where $\hat{n}_4 =\frac{1}{\sqrt{2}}\qty(1,1,0)$. This sequence therefore suppresses different noise sources hierarchically, correcting the dominant linear noise at the smallest timescale and the smaller cubic noise at a longer timescale. 
\par Another relevant choice for the axes of rotation is
\begin{equation}
    \begin{aligned}
        \hat{n}_1 &= \Big(\sqrt{\tfrac{2}{3}},0,\tfrac{1}{\sqrt{3}}\Big), & \hat{n}_3 &= \big(0,0,1\big), \\
        \hat{n}_2 &= \Big(-\tfrac{1}{\sqrt{6}},\tfrac{1}{\sqrt{2}},\tfrac{1}{\sqrt{3}}\Big), & \hat{n}_4 &= \Big(-\tfrac{1}{\sqrt{3}},0,\sqrt{\tfrac{2}{3}}\,\Big) ,
    \end{aligned}
\end{equation}
for which quadratic terms proportional to $S_z^2$ are also corrected on a smaller timescale~; the inner $\mathrm{D}_2$ sequence in this case is none other than the four-pulse sequence~\eqref{Eq.4pul}.

\par
\begin{figure}[t]
\includegraphics[width=0.95\linewidth]{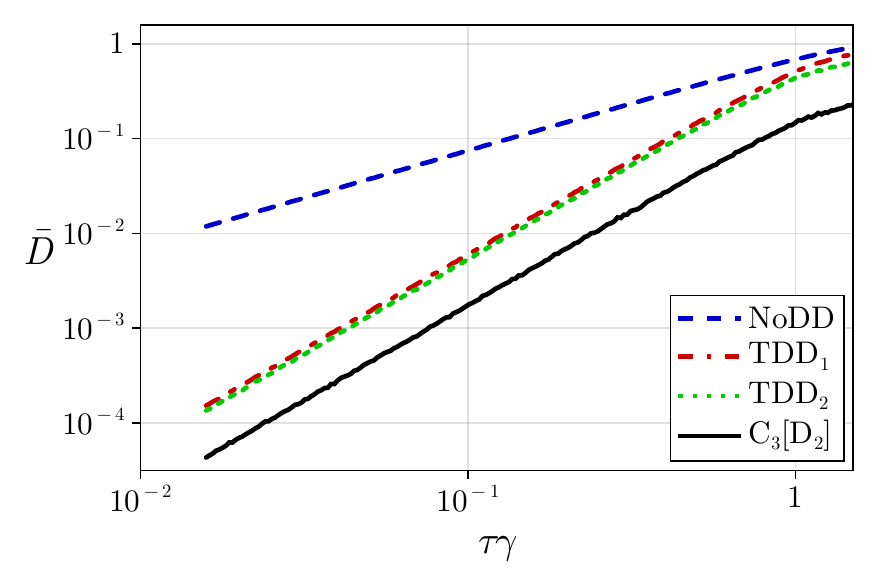}
    \caption{Distance between the identity and the noisy propagator of a single spin-$3$ with Hamiltonian $H = \gamma\, \hat{n}\bcdot\mathbf{J}$. The average distance without DD and with tetrahedral sequences $\mathrm{TDD}_1$~\eqref{TDD1}, $\mathrm{TDD}_2$~\eqref{TDD2} and $\mathrm{C}_3[\mathrm{D}_2]$~\eqref{eq.D2C3} are compared. The average is calculated on a sample of 10,000 randomly generated unit vectors $\hat{n}$.}
    \label{fig:11_distance}
\end{figure}

 As a proof of concept of this hierarchical decoupling structure, we apply the pulse sequence~\eqref{eq.D2C3} to a single spin-$j$ undergoing the coherent noise $H = \gamma\, \hat{n}\bcdot\mathbf{J}$ and calculate the distance $D(U) = \sqrt{ 1 - \frac{1}{2j+1}\abs{\tr\qty[U]}}$ between the evolution $U$ and the identity. The results for $j=3$ are plotted in Fig.~\ref{fig:11_distance} where we compare the tetrahedral sequences $\mathrm{TDD_1}$~\eqref{TDD1}, $\mathrm{TDD_2}$~\eqref{TDD2}, and $\mathrm{C}_3[ \mathrm{D}_2]$~\eqref{eq.D2C3}. The black curve corresponding to the distance for the sequence $\mathrm{C}_3[ \mathrm{D}_2]$ is consistently lower than all the others, while maintaining the same scaling with $\tau$ as the other tetrahedral sequences, thus demonstrating that the hierarchical structure provides an overall improvement in the decoupling of arbitrary linear errors.

\section{Conclusions}

In this work, we introduced a symmetry-based approach to designing dynamical decoupling sequences that are both efficient and tailored to the specific structure of the unwanted interaction Hamiltonian to be mitigated. By systematically building on the symmetries already present in the Hamiltonian, we developed a nested multi-symmetrization procedure that projects the dynamics of the system onto a subspace invariant under a chosen symmetry group. This strategy reduces the complexity of decoupling protocols and enables more targeted suppression of interactions. As a notable example, our method naturally recovers the well-known Lee-Goldburg DD sequence~\cite{LeeGoldburg_1965}, associated to the \emph{magic angle}, widely used to cancel dipole-dipole interactions in solid-state NMR. This suggests that the additional sequences we derive may likewise prove useful in NMR spectroscopy. In particular, the combination of our framework with the geometric insight offered by Majorana constellations provides an intuitive and flexible tool for constructing pulse sequences that selectively cancel interaction terms based on their symmetry properties~\cite{Levitt_2007,Mote_2016}.

\par A central result of our work is the four-pulse sequence~\eqref{Eq.4pul} that simultaneously cancels dipole-dipole interactions and on-site disorder. The eight-pulse Eulerian variant~\eqref{TEDDY} of this sequence, which we call TEDDY, was shown to be robust to finite-duration pulses as well as to many relevant control errors, including systematic over- and under-rotation as well as deviations from the desired axis of rotation and pulse shape. Furthermore, it cancels several high-order terms of the Magnus expansion in both the ideal and finite pulse regime, and we find that it outperforms state-of-the-art sequences in relevant regimes of parameters (see  Fig.~\ref{fig:6}). 

\par We have also discovered small decoupling sequences that suppress anisotropic $K$-body interactions (with $K\leq 5$) under the rotating wave approximation. These sequences have potential applications for preserving coherence in nuclear or electronic spin ensembles, as well as for probing exotic interactions between spins. Their short length, versatility, and intuitive decoupling properties make them good candidates for use as building blocks in the construction of more advanced protocols based on nesting~\cite{Lukin_2020}. We have shown that these sequences can be used as decoupling sequences for \emph{small} dephasing qudits ($d\leq 6$). Finally, we have demonstrated how the multi-symmetrization approach enables the construction of nested sequences that suppress dominant errors at shorter timescales. This is achieved by factorizing high-symmetry groups, which decouple all relevant error sources, into products of lower-symmetry subgroups, each tailored to suppress specific dominant error mechanisms. An overview of these novel DD sequences, including their description and potential use, is given in Tables~\ref{Table.4.leverage} and~\ref{Table.5.concatenation} in Appendix~\ref{App.C.summary}.

\par The same procedure of nesting and exploiting symmetry as implemented in this work could be used to achieve more efficient higher-order decoupling. In the traditional concatenation scheme for high-order decoupling~\cite{Khodjasteh_2005,Khodjasteh_2007}, a sequence is nested within another such that the outer-layer sequence essentially decouples the dominant term of the Magnus expansion left out by the inner-layer sequence. Using this framework, the outer-layer sequence can be judiciously selected to take advantage of the symmetries of the leading term of the Magnus expansion, essentially reducing the length of the high-order nested protocol. Even in the case where no apparent symmetries appear in the leading order term, knowledge of the Majorana constellation can help us identify inaccessible symmetries, providing useful information for making an informed choice about the outer-layer sequence.

\par In conclusion, we have demonstrated the importance of symmetries and their abstract decompositions in the framework of dynamical decoupling. This symmetry-based approach enables the systematic construction of efficient, robust, and unconventional DD sequences whose decoupling and robustness properties can be understood analytically, thus facilitating physical intuition. Our examples are based on point groups, where geometrical representations help elucidate the underlying symmetries of interaction subspaces and their factorizations. Although we focused on global $\mathrm{SU}(2)$ transformations, the framework extends to all symmetry groups, opening a new avenue for the development of robust protocols for dynamical decoupling and Hamiltonian engineering.

\section*{Acknowledgments}
ESE acknowledges support from the postdoctoral fellowship of the IPD-STEMA program of the University of Liège (Belgium). JM and ESE acknowledge the FWO and the F.R.S.-FNRS for their funding as part of the Excellence of Science program (EOS project 40007526). CR is a Research Fellow of the F.R.S.-FNRS. 

\section*{Data availability}
The data that support the findings of this article are openly available \cite{Colingithub}.

\appendix 
\section{Subgroup structure of the proper point groups}
\label{Ap.Subgroups}
The tetrahedral $\mathrm{T}$, octahedral $\mathrm{O}$ and icosahedral $\mathrm{I}$ exceptional point groups are equivalent to the permutation subgroups $\mathrm{A}_4$, $\mathrm{S}_4$ and $\mathrm{A}_5$, respectively~\cite{Rotman_2012}. Here, $\mathrm{S}_n$ is the permutation group of $n$ elements, and $\mathrm{A}_n$ is the alternating group of $n$ elements, that consists of the subgroup of $\mathrm{S}_n$ of even permutations (\ie, which are decomposable into an even product of cycles). We now write the maximal subgroups (\ie, a subgroup that is not contained in another proper subgroup) of each exceptional point group and their corresponding chain of subgroups~\cite{Rotman_2012,conway1985finite}
\begin{equation}
\begin{aligned}    
    \mathrm{T} \cong {}& \mathrm{A}_4 \geq \left\{ 
    \begin{array}{cc}
       \mathrm{D}_2  & \geq \mathrm{C}_2   
       \\
        \mathrm{C}_3 &  
    \end{array}
    \right.
\\[0.5cm]
    \mathrm{O} \cong {}& \mathrm{S}_4 \geq \left\{ 
    \begin{array}{rl}
       (\mathrm{T}\cong ) \mathrm{A}_4  \geq & \left\{ \begin{array}{l}
           \mathrm{D}_2 \geq \mathrm{C}_2   \\
            \mathrm{C}_3 
       \end{array} \right.  
       \\[0.5cm]
        \mathrm{D}_4 \geq & \left\{ \begin{array}{c}
            \mathrm{D}_2 \geq \mathrm{C}_2   \\
            \mathrm{C}_4 \geq \mathrm{C}_2 
       \end{array} \right.
    \\[0.5cm]
        \mathrm{D}_3 \geq & \left\{ \begin{array}{c}
            \mathrm{C}_3   \\
            \mathrm{C}_2 
       \end{array} \right.
    \end{array}
    \right.
\\[0.5cm]
    \mathrm{I} \cong {}& \mathrm{A}_5 \geq \left\{ 
    \begin{array}{rl}
    (\mathrm{T}\cong ) \mathrm{A}_4  \geq & \left\{ \begin{array}{l}
           \mathrm{D}_2 \geq \mathrm{C}_2   \\
            \mathrm{C}_3 
       \end{array} \right. 
    \\[0.5cm]
        \mathrm{D}_5 \geq & \left\{ \begin{array}{c}
            \mathrm{C}_5   \\
            \mathrm{C}_2 
       \end{array} \right.
    \\[0.5cm]
        \mathrm{D}_3 \geq & \left\{ \begin{array}{c}
            \mathrm{C}_3   \\
            \mathrm{C}_2 
       \end{array} \right.
\end{array} \right.
\end{aligned}
\end{equation}
Candidates for factorization with two complementary subgroups, $\group_1$ and $\group_2$, of a point group $\group$ are those such that $|\group|= |\group_1| |\group_2|$ and $\group_1 \cap \group_2= \{ E\}$, where $E$ is the identity element. The order of the groups is $|\mathrm{A}_n|= n!/2$, $|\mathrm{D}_n|=2n$ and $|\mathrm{C}_n|=n$. For example, the icosahedral symmetry $\mathrm{I}$ is of order $|\mathrm{I}|=|\mathrm{D}_5||\mathrm{D}_3|$. However, for all possible $\mathrm{D}_5$ and $\mathrm{D}_3$ subgroups of $\mathrm{I}$, $\mathrm{D}_5 \cap \mathrm{D}_3 \neq \{ E\} $ and consequently, $\mathrm{I}\neq \mathrm{D}_5 \mathrm{D}_3$.

For finite groups, if $\group_1 \group_2$ is also a group and $|\group_1 \group_2|=|\group_1 ||\group_2|$, then $\group_1 \group_2 = \group_2 \group_1$. Thus, any factorization is valid regardless the order.

\section{Robustness analysis of the TEDDY sequence}
\label{Ap.Robustness}
The eight-pulse sequence~\eqref{TEDDY}, written below for convenience
\begin{equation}
     \qty(\overset{\tau_0}{\mydash}~(\pi)_{\hat{n}_1} \overset{\tau_0}{\mydash}~(\pi)_{-\hat{n}_{\pm 2}})^{\cross 2} \qty(\overset{\tau_0}{\mydash}~(\pi)_{-\hat{n}_{\pm 2}}~\overset{\tau_0}{\mydash}~(\pi)_{\hat{n}_{1}})^{\cross 2},
\end{equation}
implements the quantum operation 
\begin{equation}
    S \mapsto \frac{1}{2}\sum_{\lambda=1,2}\Pi_{\mathrm{D}_2}\qty(F^{(\Gamma_{\mathrm{D}_2},\tau)}_{\lambda}(S))
\end{equation}
where 
\begin{equation}
    F^{(\Gamma_{\mathrm{D}_2},\tau)}_{\lambda}(S) = \frac{1}{\tau} \int_0^{\tau}u_{\lambda}^{\dagger}(t)\, S\, u_{\lambda}(t)dt
\end{equation}
with $u_1(\tau) = e^{-i\pi\hat{n}_1\bcdot \vec{J}}$ and $u_2(\tau) = e^{i\pi\hat{n}_{\pm 2}\bcdot \vec{J}}$. In this section, we show the robustness of the sequence to finite-duration pulses and relevant control errors. 

\subsection{Control errors}

Let us consider that the control Hamiltonian $H_1(t) = f_1(t)\hat{n}_1\bcdot \vec{J}$ implementing the pulse $u_1(t)$ comes with systematic errors which take the form of the Hamiltonian $\Delta H_1(t)$ and which is considered small compared to the time-duration of the pulse, \ie, $\tau\norm{\Delta H_1(t)} \ll 1$ $\forall t$. A relevant control error Hamiltonian would take the form 
\begin{equation}
    \Delta H_1(t) =  \Delta(t) \hat{\delta}(t)\bcdot \vec{J}\label{errorform}
\end{equation}
and would cause a deviation from the intended rotation. For instance, in the case where $\hat{\delta}(t) \equiv \hat{n}_1$, this Hamiltonian generates an over- or under-rotation. 

\par If the control error is systematic, which means that the same error term $\Delta H_1(t)$ appears every time $H_1(t)$ is turned on, the error is taken into account by changing the finite-duration term to
\begin{equation}
    \begin{aligned}
F^{(\Gamma_{\mathrm{D}_2},\tau)}_{1}(S) & \mapsto F^{(\Gamma_{\mathrm{D}_2},\tau)}_{1}(S + \Delta H_1(t)) 
\\
& =  F^{(\Gamma_{\mathrm{D}_2},\tau)}_{1}(S)+F^{(\Gamma_{\mathrm{D}_2},\tau)}_{1}(\Delta H_1(t))
\end{aligned}
\end{equation}
where 
\begin{equation}
F^{(\Gamma_{\mathrm{D}_2},\tau)}_{1}(\Delta H_1(t)) = \frac{1}{\tau} \int_0^{\tau}u_{1}^{\dagger}(t)\Delta H_1(t) u_{1}(t)dt.
\end{equation}
The finite-duration errors $F^{(\Gamma_{\mathrm{D}_2},\tau)}_{1}(S)$ and the control errors $F^{(\Gamma_{\mathrm{D}_2},\tau)}_{1}(\Delta H_1(t))$ can thus be considered individually. Crucially, if $\Delta H_1(t)$ has the form~\eqref{errorform}, then the control error Hamiltonian is corrected by the $\Pi_{\mathrm{D}_2}$ symmetrization, as $\mathrm{D}_2$ is a symmetry inaccessible to the irrep $\BHs^{(1)}$. Furthermore, when the generating pulse $u_1(t)$ has the form $e^{-i\theta_t\hat{n}_t\bcdot \vec{J}}$ at all times $t$, for some axis $\hat{n}_t$ and angle $\theta_t$ ---that is if the generating pulse represents a rotation at all times $t$---, the finite-duration error $F^{(\Gamma_{\mathrm{D}_2},\tau)}_{1}(\Delta H_1(t))$ also belongs to the irrep $\BHs^{(1)}$, as it is a closed subspace under $\mathrm{SU}(2)$ unitary transformation. Consequently, it is suppressed by the symmetrization $\Pi_{\mathrm{D}_2}$, which proves robustness to such control errors to the first order of the Magnus expansion.
\par Evidently, the same argument can be made for the second generating pulse $u_2(t)$. 
\subsection{Finite-duration errors}

Even when $S$ belongs to the so-called \emph{correctable subspace} of the $\mathrm{D}_2$ group, this does not guarantee that the finite duration error $F^{(\Gamma_{\mathrm{D}_2},\tau)}_{1}(S)$ is corrected by $\mathrm{D}_2$. However, in the case where 
\begin{equation}
    u_1(\tau) = e^{-i\int_0^\tau f_1(t)dt \, \hat{n}_1\bcdot \vec{J}},
\end{equation}
and if $f_1(t) = f_1(\tau - t)$, we can show that 
\begin{equation}\label{PiD2equality}
\Pi_{\mathrm{D}_2}\qty[\int_0^{\tau}u_1^{\dagger}(t)Su_1(t)dt]= \int_0^{\tau}u_1(t)\Pi_{\mathrm{D}_2}\qty[S]u_1^{\dagger}(t)dt.
\end{equation}
This equality means that, if $S$ is corrected by $\Pi_{\mathrm{D}_2}$, then $F^{(\Gamma_{\mathrm{D}_2},\tau)}_{1}(S)$ is also corrected. An analogous argument applies to the second generating pulse, $u_2(\tau)$.

In what follows, we set $\tilde{S}_1(t)=u_1^{\dagger}(t)Su_1(t)$. The proof of the equality \eqref{PiD2equality} begins by writing
\begin{widetext}
    \begin{equation}\begin{aligned}
\Pi_{\mathrm{D}_2}\qty[\int_0^{\tau}\tilde{S}_1(t)dt] ={}& \int_0^{\tau}\tilde{S}_1(t) dt
    + e^{i\pi\hat{n}_1\bcdot \vec{J}}\left(\int_0^{\tau}\tilde{S}_1(t)dt\right) e^{-i\pi\hat{n}_1\bcdot \vec{J}} \\
    &+ e^{i\pi\hat{n}_{+2}\bcdot \vec{J}}\left(\int_0^{\tau}\tilde{S}_1(t) dt\right) e^{-i\pi\hat{n}_{+2}\bcdot \vec{J}}
    + e^{i\pi\hat{n}_{-2}\bcdot \vec{J}}\left(\int_0^{\tau}\tilde{S}_1(t) dt\right) e^{-i\pi\hat{n}_{-2}\bcdot \vec{J}}
\end{aligned}\label{bigequation}\end{equation}
\end{widetext}
and realizing that since $\hat{n}_{\pm 2}$ are orthogonal to $\hat{n}_1$, we have 
\begin{equation}\begin{aligned}
    &e^{i\pi\hat{n}_{\pm2}\bcdot \vec{J}}\left(\int_0^{\tau}\tilde{S}_1(t)dt\right)e^{-i\pi\hat{n}_{\pm2}\bcdot \vec{J}} \\&= \int_0^{\tau}u_1(t)e^{i\pi\hat{n}_{\pm2}\bcdot \vec{J}}Se^{-i\pi\hat{n}_{\pm2}\bcdot \vec{J}}u^{\dagger}_1(t)dt.
\end{aligned}\end{equation}

Additionally, because of the symmetric control profile, we find that 
\begin{equation}
    u_1(t) = e^{-i\pi\hat{n}_{1}\bcdot \vec{J}}e^{i\int_0^{\tau -t}f(t')dt'\, \hat{n}_{1}\bcdot \vec{J}}
\end{equation}
which can be used to prove that 
\begin{equation}
    \int_0^{\tau}\tilde{S}_1(t)dt = \int_0^{\tau}u_1(t)e^{-i\pi\hat{n}_{1}\bcdot \vec{J}}Se^{i\pi\hat{n}_{1}\bcdot \vec{J}}u_1^{\dagger}(t)dt.
\end{equation}
Putting everything together,~\eqref{bigequation} becomes 
\begin{widetext}\begin{equation}\begin{aligned}\Pi_{\mathrm{D}_2}\qty[\int_0^{\tau}\tilde{S}_1(t)dt] &= \int_0^{\tau}u_1(t)e^{-i\pi\hat{n}_1\bcdot \vec{J}}Se^{i\pi\hat{n}_1\bcdot \vec{J}}u^{\dagger}_1(t)dt
    + \int_0^{\tau}u_1(t)Su_1^{\dagger}(t)dt \\
    &+ \int_0^{\tau}u_1(t)e^{i\pi\hat{n}_{+2}\bcdot \vec{J}}Se^{-i\pi\hat{n}_{+2}\bcdot \vec{J}}u_1^{\dagger}(t)e^{-i\pi\hat{n}_{+2}\bcdot \vec{J}}dt
    +\int_0^{\tau}u_1(t) e^{i\pi\hat{n}_{-2}\bcdot \vec{J}}S e^{-i\pi\hat{n}_{-2}\bcdot \vec{J}}u_1^{\dagger}(t)dt \\
    &= \int_0^{\tau}u_1(t)\Pi_{\mathrm{D}_2}\qty[S]u_1^{\dagger}(t)dt.
\end{aligned}\label{bigequation2}\end{equation}
\end{widetext}
This demonstrates that the sequence is robust against finite-duration pulse errors under moderate conditions in terms of pulse profile.

\section{Overview of DD sequences}
\label{App.C.summary}
Tables~\ref{Table.4.leverage} and \ref{Table.5.concatenation} summarise the most relevant DD sequences developed in the work, which exploit existing symmetries in the interaction Hamiltonian and hierarchical decoupling, respectively.
\renewcommand{\arraystretch}{1.3}
\setlength{\arrayrulewidth}{0.5pt}
\begin{table*}[t]
    \centering
    \begin{tabular}{|c|c|c|c|}
    \hline 
      \multirow{2}{*}{Description} & \# of & \multirow{2}{*}{Pulse  sequence}  & \multirow{2}{*}{Rotation axes} 
      \\    &  pulses &  & 
      \\ \hline \hline
         \multirow{2}{*}{\centering \begin{tabular}{c}
          DD sequence \eqref{Eq.Seq.3P} for $H_{\mathrm{dd}}^{\mathrm{RWA}}$ \\
          c.f.\ Lee-Goldburg sequence~\cite{LeeGoldburg_1965} 
     \end{tabular}
     }  
         & \multirow{2}{*}{3} & \multirow{2}{*}{$ \left( \overset{\tau_0}{\mydash}~\qty(\frac{2\pi}{3})_{\hat{n}} \right)^{\times 3}$ } & \multirow{2}{*}{$\hat{n} = \frac{1}{\sqrt{3}}\qty(-1,1,1)$}  
    \\[2pt] & & & 
    \\[2pt]  \hline \hline
         \multirow{2}{*}{DD sequence \eqref{Eq.4pul} for $H_{\mathrm{dd}+\mathrm{dis}}^{\mathrm{RWA}}$} 
         & \multirow{2}{*}{4} &
         \multirow{2}{*}{$ \left( \overset{\tau_0}{\mydash}~(\pi)_{\hat{n}_1}~\overset{\tau_0}{\mydash}~(\pi)_{\hat{n}_{\pm 2}} \right)^{\times 2}$ } & 
    \\[2pt] &  & & 
    \multirow{6}{*}{ \begin{tabular}{rl}
    $\hat{n}_1 =$ & $ \frac{1}{\sqrt{3}}\qty(\sqrt{2},0,1)$
    \\[6pt]
         $\hat{n}_{\pm 2} =$& $\qty(-\frac{1}{\sqrt{6}}, \pm\frac{1}{\sqrt{2}}, \frac{1}{\sqrt{3}})$ 
    \end{tabular}}
    \\[4pt] \cline{1-3}
     \multirow{2}{*}{\begin{tabular}{c}
          TEDDY sequence \eqref{TEDDY} for $H_{\mathrm{dd}+\mathrm{dis}}^{\mathrm{RWA}}$  \\
          robust to systematic control errors 
     \end{tabular}}  
          & \multirow{2}{*}{8} & \multirow{2}{*}{$     \qty(\overset{\tau_0}{\mydash}\,(\pi)_{\hat{n}_1}\,\overset{\tau_0}{\mydash}\,(\pi)_{\hat{n}_{\pm 2}})^{\cross 2} \qty(\overset{\tau_0}{\mydash}\,(\pi)_{\hat{n}_{\pm 2}}\,\overset{\tau_0}{\mydash}\,(\pi)_{\hat{n}_{1}})^{\cross 2}$}  & 
    \\[2pt]
    & & &  
    \\[4pt] \cline{1-3}
        $(\mathrm{TEDDY})(\mathrm{TEDDY})^{\dagger}$   & \multirow{3}{*}{16} & \multirow{3}{*}{$(\mathrm{TEDDY})(\mathrm{TEDDY})^{\dagger}$} & 
    \\[2pt] 
     2nd order decoupling for $H_{\mathrm{dd}+\mathrm{dis}}^{\mathrm{RWA}}$ & & &
    \\[2pt] 
        robust to systematic control errors  & & &
    \\[2pt] \hline \hline
\multirow{4}{*}{\begin{tabular}{c}
     DD sequences \eqref{eq:D3a} and \eqref{eq:D3b} for\\ 3-body
      multilinear int.\ under RWA \\
      (dephasing Hamiltonians)
\end{tabular}}
          & \multirow{2}{*}{6}  & \multirow{2}{*}{$\qty(\overset{\tau_0}{\mydash}~(\pi)_{\hat{n}_1}~ \overset{\tau_0}{\mydash}~(\pi)_{\hat{n}_2} )^{\cross 3}$ } & 
          \multirow{4}{*}{\begin{tabular}{rl}
               $\hat{n}_1=$& $\frac{1}{\sqrt{2}}\qty(1,-1,0)$  \\[5pt]
               $\hat{n}_2=$& $\frac{1}{\sqrt{2}}\qty(0,-1,1)$ \\[5pt]
               $\hat{n}_3=$& $\frac{1}{\sqrt{3}}\qty(1,1,1)$
          \end{tabular}}
    \\[2pt]  & & &
    \\[2pt] \cline{2-3}  & \multirow{2}{*}{6} &     \multirow{2}{*}{$ \qty(\overset{\tau_0}{\mydash}~\qty(\frac{2\pi}{3})_{\hat{n}_3}~\overset{\tau_0}{\mydash}~\qty(\frac{2\pi}{3})_{\hat{n}_3}~\overset{\tau_0}{\mydash}~(\pi)_{\hat{n}_1})^{\cross 2}$}
 & 
    \\[2pt] & & &
    \\[2pt] \hline \hline
\multirow{4}{*}{\begin{tabular}{c}
     DD sequences for 4 and 5-body \\
      multilinear interactions under RWA \\
      (dephasing Hamiltonians)
\end{tabular}}
          & \multirow{2}{*}{12}  & \multirow{2}{*}{See Eq.~\eqref{TDD1}} & \multirow{4}{*}{See Eqs.~\eqref{Eq.Axes.4.5} and \eqref{Eq.Axes.4.5.body}}
    \\[2pt]  & & &
    \\[2pt] \cline{2-3} & \multirow{2}{*}{12}  & \multirow{2}{*}{See Eq.~\eqref{TDD2}} & 
    \\[2pt] & &  & 
    \\[2pt]
         \hline     
    \end{tabular}
    \caption{Summary of DD sequences that leverage existing symmetries in the interaction Hamiltonian. The $H_{\mathrm{dd}}^{\mathrm{RWA}}$~\eqref{eq:Dip.Ham.} ($H_{\mathrm{dd}+\mathrm{dis}}^{\mathrm{RWA}}$~\eqref{eq:Dip.Dis.Ham.}) Hamiltonian considers dipole-dipole ($+$ disorder) interactions under the rotating wave approximation (RWA).}
    \label{Table.4.leverage}
\end{table*}

\begin{table*}[t]
    \centering
    \begin{tabular}{|c|c|c|c|}
    \hline 
      \multirow{2}{*}{Description} & \# of & \multirow{2}{*}{Pulse  sequence}  & \multirow{2}{*}{Rotation axes} 
      \\    &  pulses &  & 
      \\ \hline \hline
         $\mathrm{T}= \mathrm{C}_3[\mathrm{D}_2]$ sequence \eqref{eq.D2C3}
         & \multirow{3}{*}{12} & \multirow{3}{*}{$\qty( \overset{\tau_0}{\mydash}~(\pi)_{\hat{n}_1}~\overset{\tau_0}{\mydash}~(\pi)_{\hat{n}_{ 2}}~\overset{\tau_0}{\mydash}~(\pi)_{\hat{n}_1}~\overset{\tau_0}{\mydash}~(\pi)_{\hat{n}_{2}}~\qty(\frac{2\pi}{3})_{\hat{n}_3})^{\cross 3}$ } & \multirow{6}{*}{\begin{tabular}{rl}
               $\hat{n}_1=$&$(1,0,0)$ \\ $\hat{n}_2=$&$(0,1,0)$ \\ $\hat{n}_3=$&$\frac{1}{\sqrt{3}}\qty(1,1,1)$ \\
               $\hat{n}_4=$&$\frac{1}{\sqrt{2}}\qty(1,1,0)$
         \end{tabular}}  
    \\[2pt] primarily suppresses linear errors & & & 
    \\[2pt] and secondarily quadratic errors & & & 
    \\[2pt]  \cline{1-3}  
      $\mathrm{O}= \mathrm{C}_2 [\mathrm{C}_3[\mathrm{D}_2]]$ sequence \eqref{Eq.O.hierarchy} & \multirow{3}{*}{24}& 
      \multirow{3}{*}{$\qty(\mathrm{C}_3\qty[\mathrm{D}_2]\,(\pi)_{\hat{n}_4})^{\cross 2}$}&  
    \\[2pt] primarily suppresses linear errors, then & & & 
    \\[2pt] quadratic and minimally cubic errors & & & 
    \\[2pt]
         \hline     
    \end{tabular}
    \caption{Summary of DD sequences that decouple several interactions hierarchically.}
    \label{Table.5.concatenation}
\end{table*}
\newpage 
\bibliographystyle{apsrev4-2}
\bibliography{References}
\end{document}